**Estimates of dissipation of wave energy by sea ice for a field experiment in the Southern Ocean, using model/data inversion**

W. Erick Rogers, Naval Research Laboratory, Stennis Space Center, MS, USA

Michael H. Meylan, The University of Newcastle, Australia

Alison L. Kohout, National Institute of Water and Atmospheric Research (NIWA), New Zealand

Corresponding author: W. Erick Rogers (erick.rogers@nrlssc.navy.mil)

*Keywords*

Ocean waves; sea ice; spectral wave model; wave-ice interaction; WAVEWATCH III; dissipation by sea ice

*Highlights*

- Frequency dependence of dissipation by sea ice is estimated for a very large dataset.

- Power dependence on frequency is found to be power two to four.

- A clear negative correlation is found between waveheight and dissipation rate.

- Correlation of dissipation with ice thickness can be exploited in predictive models.

*Long-form highlights*

- Frequency dependence of dissipation by sea ice is estimated for a very large dataset (9477 records) near Antarctica.

- Power dependence of dissipation rate on frequency is found to be most predominantly either a power two and three binomial or a power two and four binomial.

- At frequencies in the energy-containing region of the spectrum, there is a clear negative correlation between waveheight and dissipation rate, though causality is unknown.

- In the energy-containing region of the spectrum, we find the expected positive correlation between ice thickness and dissipation rate, which can potentially be exploited for predictive models.

**Abstract**

A model-data inversion is applied to a very large observational dataset collected in the Southern Ocean north of the Ross Sea during late autumn to early winter, producing estimates of the frequency-dependent rate of dissipation by sea ice. The modeling platform is WAVEWATCH III® which accounts for non-stationarity, advection, wave generation, and other relevant processes. The resulting 9477 dissipation profiles are co-located with other variables such as ice thickness to quantify correlations which might be exploited in later studies to improve predictions. Mean dissipation profiles from the inversion are fitted to simple binomials. Variability about the mean profile is not small, but the binomials show remarkable qualitative similarity to prior observation-based estimates of dissipation, and the power dependence is consistent with at least three theoretical models, one of which assumes that dissipation is dominated by turbulence generated by shear at the ice-water interface.

## 1. Introduction

During the past 20 years, open-ocean spectral wave modeling has achieved high levels of accuracy, e.g. Bidlot (2018), largely driven by improvements to accuracy in wind forcing and physics parameterizations (e.g. Ardhuin et al. 2010, Liu et al. 2019). However, areas near coastlines and the sea ice margins remain a challenge, with relatively large errors and therefore substantial improvements possible (e.g. Cavaleri et al. 2018 and Sutherland et al. 2018, respectively). Predictions near the ice edge fall into two broad categories: the off-ice wind scenario, and the on-ice wave scenario. In the former case, errors are caused by uncertainty in the ice edge location, and challenges with describing the rarely-studied conditions of growth in partial ice cover and the ubiquitous atmospheric instability (i.e. cold air blowing from ice onto warmer water), e.g. Gemmrich et al. (2018). Modeling the on-ice wave scenario also suffers from uncertainty of the ice edge position. In addition, wave conditions within the ice are dominantly determined by the model's representation of dissipation by sea ice. Fundamentally, this dissipation is strongly dependent on frequency (e.g. Wadhams et al. 1988). The objective of this paper is to use observational data to estimate this frequency-dependent dissipation rate and study possible dependencies, to guide future improvements to parameterizations of dissipation used by spectral wave models, and thence improvements to accuracy of these models in on-ice wave conditions.

More recently, there has been a surge in research activity on the topic of wave-ice interaction. This includes theoretical studies, numerical modeling studies, laboratory studies, and studies using field observations, either in situ observations or those from remote sensing. To demonstrate this quantitatively, we surveyed peer-reviewed articles published during 2015 to 2019. We found 33 articles for which the dissipation of wave energy by sea ice is a primary feature (e.g. Toffoli et al. 2015, Ardhuin et al. 2016, Asplin et al. 2018, Boutin et al. 2018, Herman et al. 2019, Marchenko et al. 2019, Voermans et al. 2019, Yue et al. 2019). In addition, we find nine other articles which are primarily concerned with both dissipation and changes to wavelength caused by the ice (e.g. Li et al. 2015b, Zhao et al. 2017), and four that are primarily concerned with the changes to wavelength (e.g. Sree et a. 2018). We find nine other articles primarily about the scattering and reflection of wave energy by sea ice (e.g. Montiel et al. 2016, Orzech et al. 2016, Bennetts and Williams 2015). We find 15 articles which treat the impact of waves on ice, either by fracturing the ice (e.g. Bennetts et al. 2017), or by determining the ice type during growth (Roach et al. 2018, Shi et al. 2019). We find seven review articles during this period (e.g. Shen 2019, Squire 2019). Detailed review of all these topics is beyond the scope of this paper, but the latter two review articles are a good starting point for interested readers.

Among wave-ice interaction studies, observational campaigns in the southern hemisphere are in the minority. Perhaps the earliest study was that of Robin (1963) describing measurements using a ship-borne wave recorder in the Weddell Sea during 1959-1960. These measurements used early technology, and since wave observations were local to the ship, results were described not in terms of dissipation rate (the key parameter required by a modern numerical wave model), but instead in terms of the distance of penetration of swells into the ice field. They found that this value depends on three rather intuitive parameters: wave period (or length), ice thickness, and floe size, with the first being important in all cases, the second being important for longer waves, and the third being important for shorter waves. Doble and Wadhams (2006) and Doble (2009) describe a buoy deployment in the Weddell Sea in 2000, with particular attention to autumn ice growth and interaction with pancake and frazil ice. Doble and Bidlot (2013) study data from a buoy from the same deployment, but five months later (late winter/early spring), when the buoy was in pack ice during an energetic breakup event. Kohout et al. (2014) and Meylan et al. (2014) analyzed data from a buoy deployment during the SIPEX II voyage in 2012 north of Wilkes Land, Antarctica; these data have since been used also by Li et al. (2015a) and Liu et al. (2020). In July 2017, C. Eayrs (New York Univ.) and others deployed wave buoys north of Queen Maud Land, Antarctica (Vichi et al. 2019); Alberello et al. (2019) study pancake and frazil ice formation during the same experiment.

The present study uses another buoy dataset collected only a month earlier (June 2017), part of a larger dataset collected north of the Ross Sea. This builds on a previous analysis by Kohout et al. (2020), who looked at the decay rate of total energy (i.e. significant waveheight) within the sea ice for the full wave buoy dataset (April to July 2017). That study found that the decay is generally linear exponential, and concluded that dissipation increases with ice concentration. The present study uses model-data inversion to estimate the frequency distribution of dissipation rate. A notable advantage of the model-data inversion method is that it does not require that wave energy travels primarily along an axis between two buoys.

A remarkable feature of the Kohout et al. (2020) dataset is its size: we are not aware of any prior study of spectral dissipation by sea ice using a larger dataset. The full dataset includes over 14,000 spectra (April 21 to July 26), and the present study uses 9477 of those records for co-locations during June 6 to 30. This large population enables robust evaluation of correlation between the dissipation rate and several additional variables ("tertiary variables"), such as satellite-based ice thickness estimates.

The paper is structured as follows. Section 2 reviews the literature on the topic of the frequency dependence of dissipation. Section 3 gives a brief overview of the field experiment. Section 4 describes the wave model, Section 5 describes the model-data inversion, and Section 6 presents results of the analysis. Section 7 presents possible simple fits to the mean dissipation profiles. Sections 8 and 9 contain discussion and conclusions.

## 2. Brief review of frequency dependence in the literature

Most observations suggest that the $S_{ice}$ source term acts as a low-pass filter. Contrary observations of a "roll-over" effect, i.e. lower dissipation rate at high frequencies relative to mid-frequencies, have long been suspected as an artifact of disregarding local windsea growth (Wadhams et al. 1988; Rogers et al. 2016; Li et al. 2017). Empirical parametric representation of the low-pass filter should logically be determined by the overall shape of $k_i(f)$. If the $k_i(f)$ profile levels off, then a tanh or error function may be most suitable. If the $k_i(f)$ profile increases continuously within the wave model's prognostic frequency range (typically ending at 0.5 to 1.0 Hz), then a power dependence, $k_i(f) \propto f^n$ may be suitable, and is in fact popular in the literature, perhaps since many proposed theoretical relations reduce to this form. Here, we summarize values of power dependence $n$ which we have found in the literature.

- ❑ Fit to data derived from sonar observations in Greenland Sea, Wadhams (1978), according to Squire 1998, $n$=2
- ❑ A viscoelastic model, the first of two new models proposed by Meylan et al. (2018), $n$=2
- ❑ A viscoelastic model, the second of two new models proposed by Meylan et al. (2018), $n$=3
- ❑ Derived from buoy observations near Antarctica, Meylan et al. (2014), binomial with $n_1$=2 and $n_2$=4. Meylan et al. (2018) report a best fit of $n$=1.9 for the same dataset.
- ❑ Derived from buoy observations in pancake ice near Antarctica, Doble et al. (2015): $n$=2.1. Meylan et al. (2018) report a best fit of $n$=2.9 for the same dataset. Sutherland et al. (2019) report a best fit of $n$=3.8 for this dataset, using a subset of the total frequency range.
- ❑ Derived from observations in the Greenland and Bering Seas, reported in the seminal work of Wadhams et al. (1988). This dataset is highly scattered, either due to highly varied conditions, or less precise methods. In Meylan et al. (2018), a best fit of $n$=3.6 is reported for this dataset.
- ❑ Derived from buoy observations in/near Beaufort Sea, Rogers et al. (2016, 2018a), binomial with $n_1$=2 and $n_2$=4 to 5.
- ❑ For the same experiment in the Beaufort Sea, Cheng et al. (2017) estimated dissipation rates, and Meylan et al. (2018) report that a best fit of $n$=3.6 (matching their best-fit $n$ for Wadhams et al. (1988)). It is shown in Supplemental Information that superb fit can also be found with the Cheng dataset using binomial with $n_1$=3 and $n_2$=4.
- ❑ Viscoelastic model of Robinson and Palmer (1990) as reduced by Meylan et al. (2018), $n$=3. See also Liu et al. (2020).
- ❑ Boundary layer model, by F. Ardhuin (newer variant of 'IC2' in WAVEWATCH III): turbulent regime: $n$=3 and laminar regime: $n$=3.5. This is described in Appendix B of Stopa et al. (2016).
- ❑ Inextensible surface cover, Weber (1987): $n$=3.5. This is treated in Lamb (1932) under the heading "calming effect of oil on waves", and is based on the idea that horizontal motion is prevented in the ice but not the water, so a boundary layer develops at the interface. Weber (1987) invokes the concept of eddy viscosity to represent turbulence in the lower/water layer. Liu et al. (1991) also represent dissipation by turbulence generated at the ice-water interface using this eddy viscosity concept, and this was the original form of the dissipation term "IC2" in WAVEWATCH III as implemented by Rogers and Orzech (2013). When the group velocity is proportional to wave period (as in deep, open water), and if ice inertial effects are disregarded, the Liu et al. (1991) model gives $n$=3.5.

- ❑ Empirical fit to airborne data by Sutherland et al. (2018): $n$=3.5
- ❑ Friction model, Kohout et al. (2011): $n$=4
- ❑ Viscous water model, Lamb (1932), Weber (1987): $n$=5
- ❑ Empirical fit to laboratory data by Rabault et al. (2019): $n$=6 to 6.5
- ❑ Viscous ice model, Keller (1998) (this can be recovered in WAVEWATCH III using the 'IC3' viscoelastic model with zero elasticity): $n$=7
- ❑ Viscoelastic model of Mosig et al. (2015). This is 'extended Fox and Squire model' and is related to model of Greenhill (1887), see Meylan et al. (2018), $n$=11.

More detailed descriptions of some of these can be found in Meylan et al. (2018) and Rogers et al. (2018a,b).

## 3. Field experiment (overview)

The wave observations used in this study are described in Kohout and Williams (2019) and Kohout et al. (2020). Here, we provide a brief overview.

During the Polynyas, Ice Production, and seasonal Evolution in the Ross Sea (PIPERS) expedition in 2017, 14 waves-in-ice observation systems (WIIOS, Kohout et al. 2015) were deployed on Antarctic sea ice, in the Southern Ocean north of the Ross Sea. Four buoys were deployed during the transit south toward the Ross Sea on 21-22 April 2017, and ten more were deployed during the outbound transit, 30 May to 3 June, with the first data recorded on 2 June. The inbound deployment was near 172˚E and the outbound deployment was approximately 470 km east of that, near 184˚E. Buoy survival durations varied significantly, with the last record of the western (inbound) group coming on 6 July (10.8 week duration) and the last record of the eastern (outbound) group on 26 July (7.7 week duration). Figure 1 shows the geographic location of the PIPERS deployment relative to three other notable wave-ice field experiments near Antarctica: an experiment in 2000 in the Weddell sea reported by Doble and Wadhams (2006); an experiment in 2012 north of Wilkes Land reported in Kohout et al. (2014) and Meylan et al. (2014); and an experiment in July 2017 with data collected by C. Eayrs, north of Queen Maud Land, reported in Vichi et al. (2019).

One buoy in the eastern group had a GPS failure, so effectively there are four and nine buoys in the west and east group, respectively. The spectral energy, $E(f)$, data are provided with a precision of $1 \times 10^{-6}$ m²/Hz. The full dataset includes 23,206 spectra with valid GPS coordinates. Of these, records with significant waveheight $H_{m0}$<0.1 m are omitted here to improve overall signal-to-noise ratio, leaving 14,602 spectra. In this study, we analyze data from the eastern (outbound) grouping during a 24-day period of the transition from late autumn to early winter: 6 to 30 June. This subset includes 9511 spectra and all of the large wave events (observed $H_{m0}$>3 m) of the full dataset. As will be described in Section 5, the model-data inversion requires us to omit cases of low ice concentration, leaving 9477 spectra. The final number represents 41% of the full dataset, and is at least 35 times larger than the dataset collected in SIPEX II (with 268 records), as previously noted by Kohout and Williams (2019).

Six of the buoys in the eastern group have more than 500 records each; information about these buoys are described in Table 1, and time series of the distance to the ice edge[1] for each of these six buoys is shown in Figure 2. Each wave buoy is inside a Pelican case deployed directly on the ice, either on an existing floe, or on continuous ice, with the expectation that it will be on a floe subsequent to an ice breakup event. The buoys are designed to never be retrieved, so all processing is performed on board and data is sent by satellite telemetry. The primary component of the buoy is an Inertial Measurement Unit (IMU). Two of the nine buoys in the eastern grouping deployed deepest in the ice used high precision IMUs, with the expectation that they would experience the lowest wave energy; these included buoy 14, listed in Table 1. The buoys were tested using a mechanical wave simulator in the laboratory. One was also used in a field test in open water with light wave conditions (peak frequency $f_p$=0.4 to 0.5 Hz) and demonstrated good high-frequency tail shape of the spectrum $E(f)$ as low as $E(f) = 2 \times 10^{-6}$ m²/Hz and good agreement with another buoy deployed for comparison (Thomson 2012). These tests are described in Kohout and Williams (2019).

The WIIOS buoys record the motion of the floe upon which they rest. They can be expected to measure the motion of the underlying sea surface if either the floe is small relative to the wavelength being measured, or if the floe is large but bending freely such that it follows the sea surface. In either case, the response function is 1:1. However, if the floe is large and rigid, there will be deviation, particularly for higher frequencies. This typically means that the motion is damped, i.e. measured $E(f)$ of the buoy is less than unmeasured $E(f)$ of the sea surface, e.g. see Thomson et al. (2015) for the relevant formula and Steele et al. (1985, Table 1) for an example of this "vertical displacement Response Amplitude Operator (RAO)" for a 10-m diameter aluminum discus buoy.

In this region and season, southerly katabatic winds push new ice northward from the Ross Sea (Kohout and Williams 2019) while wave conditions are dominated by extratropical storms in the Southern Ocean which generate swells arriving primarily from the northwest and (less commonly) north. The waves may transfer momentum to the mean flow and ice as they decay (Liu and Mollo-Christensen 1988; Stopa et al. 2018), which would result in a southward stress countering the wind stress. The eastern buoys were deployed along a south to north (meridional) transect, with the strategy of measuring dissipation of swells traveling from a northerly direction, but they drifted to a roughly west-southwest to east-northeast orientation by mid-June. (Diagrams and an animation of the buoy tracks are provided with the Supplemental Information.) Ship-based ice observations were made using Antarctic Sea Ice Processes and Climate (ASPeCt) protocol (Worby 1999), and cores were taken during some buoy deployments. Ice consisted primarily of floes from new sheet ice (15-30 cm thick) and first year ice (30 to 70 cm thick). This is summarized in Table 1 and more details can be found in Kohout et al. (2020).

Figure 3 shows an example of wave spectra for the six buoys, using the same color scheme as used in Figure 2. By comparing the two figures, one can see the correspondence between the damping of the spectrum and distance from the ice edge, with damping occurring first at higher frequencies, and then more noticeably at the dominant frequencies for the buoys further in the ice. At the lowest energy levels, the high frequency tail slope is flatter than what might be expected in a spectrum in ice, recalling that ice acts as a low-pass filter, which should result is a steep slope. The tail may be "propped up" by an unknown physical process, but it is also credible that the feature could be simply caused by instrument noise. The energy levels below which the spectral tails are noticeably elevated—either by noise or the unknown physical process—are indicated in Figure 3. This is discussed further in Section 8.

---

[1] see Section 6.1 for definition

Table 1. WIIOS deployments for PIPERS-17: Only buoys in eastern grouping with 500 or more data records are included. Notation: $h_{ice}$, $d_{ice}$, $a_{ice}$ are ice thickness, floe size, and concentration respectively. Buoys are given here in order of deployment, from south to north. Note that the sizes of the floes on which the buoys are deployed are not known generally. Floes may have broken after deployment, and some buoys far from the ice edge were actually deployed on the continuous ice, and became "buoys on floes" later.

| Experi-ment | | | floe upon which the buoy is deployed | | most prevalent ice near the buoy, from nearest ASPeCt record | | | # of spectra | init. dist. from ice edge (km) |
|---|---|---|---|---|---|---|---|---|---|
| | Buoy # | Buoy ID | $h_{ice}$ (cm) | $d_{ice}$ (m) | $a_{ice}$ | $h_{ice}$ (cm) | $d_{ice}$ (m) | | |
| PIPERS-17 | 14 | A-34* | 54 | N/A (cont. ice) | 100% | 50 | 100-500 | 509 | 244 |
| | 5 | B-25 | 60 | 100 | 100% | 60, 75 | 20-100 | 1349 | 175 |
| | 6 | B-26 | 70 | 20 | 100% | 30 | 20-100 | 1746 | 153 |
| | 7 | B-27 | 36 | 40 | 50% | 30 | 20-100 | 1830 | 151 |
| | 9 | B-29 | 50 | 40 | 100% | 20 | <20 (cake) | 1668 | 133 |
| | 10 | B-30 | 75 | 20 | 100% | 30 | 20-100 | 2052 | 118 |

* equipped with high-precision "Kistler" accelerometer

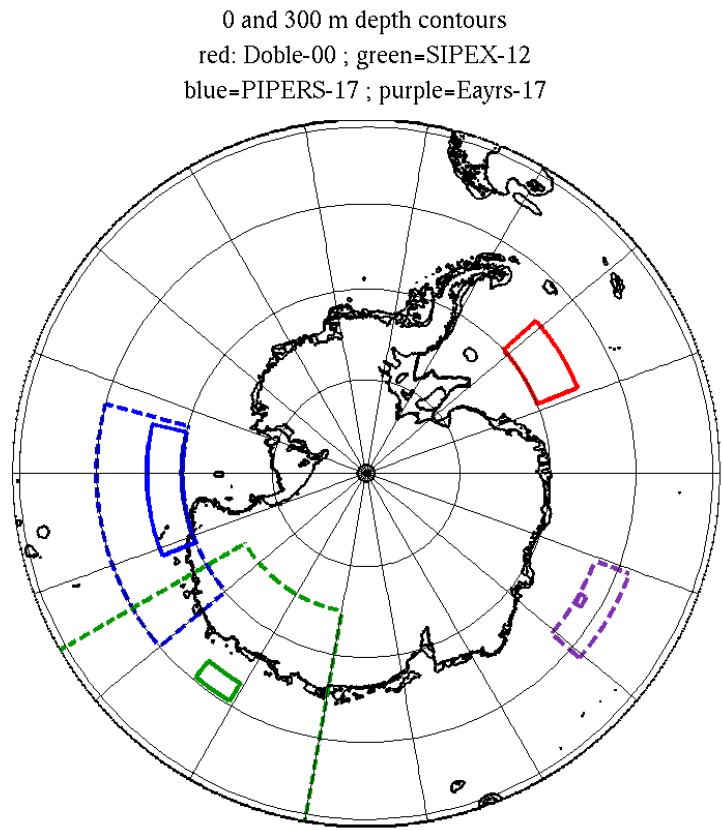

0 and 300 m depth contours
red: Doble-00 ; green=SIPEX-12
blue=PIPERS-17 ; purple=Eayrs-17

Figure 1. Geographic location of the observational campaign used in this paper (PIPERS-17) is marked in blue. Other observational studies are indicated as historical context and potential further work with comparable inversion exercises. Dashed colored lines are model grids and solid colored lines approximately denote the bounds of observational data for each case. Dashed blue box (for PIPERS-17) is the model domain of this study. Dashed green box (for SIPEX-12) is from Li et al. (2015a) (1/4° resolution). Dashed purple box (for Eayrs-17) pertains to a hindcast that will be reported on separately.

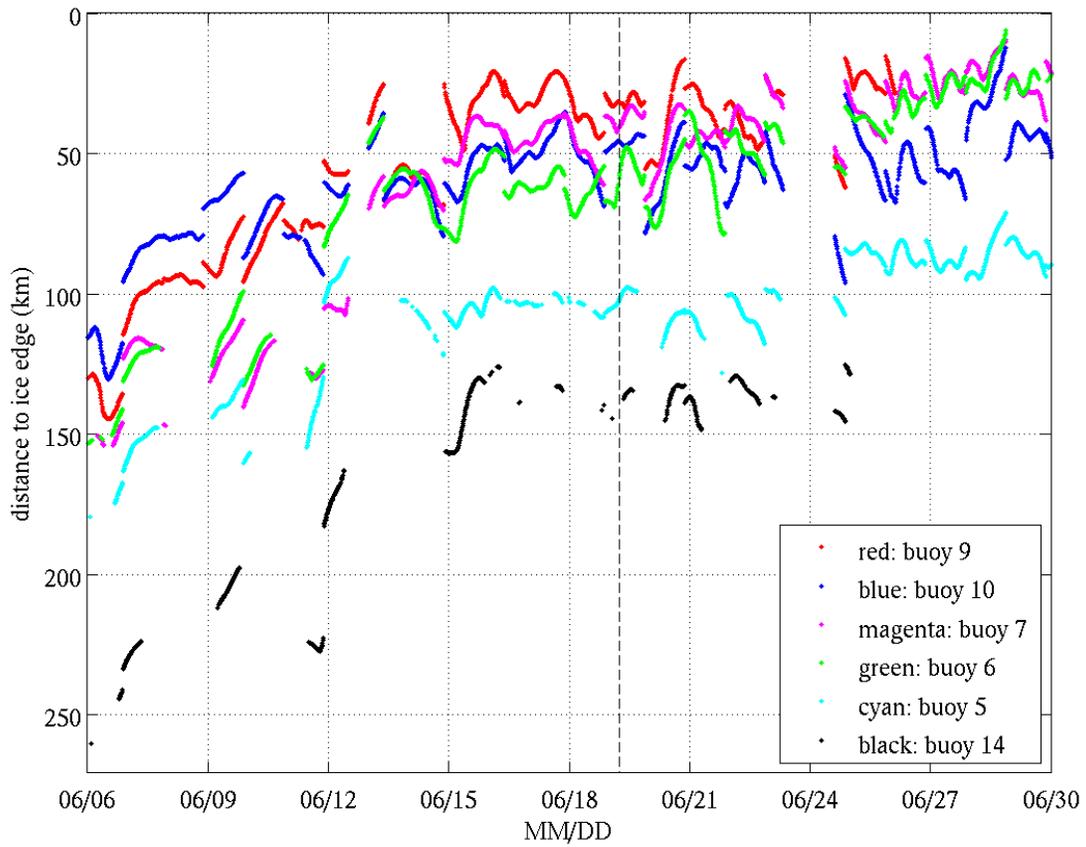

Figure 2. Positions of the six buoys, relative to ice edge. The vertical line here indicates the time used in Figures 3, 4, and 5.

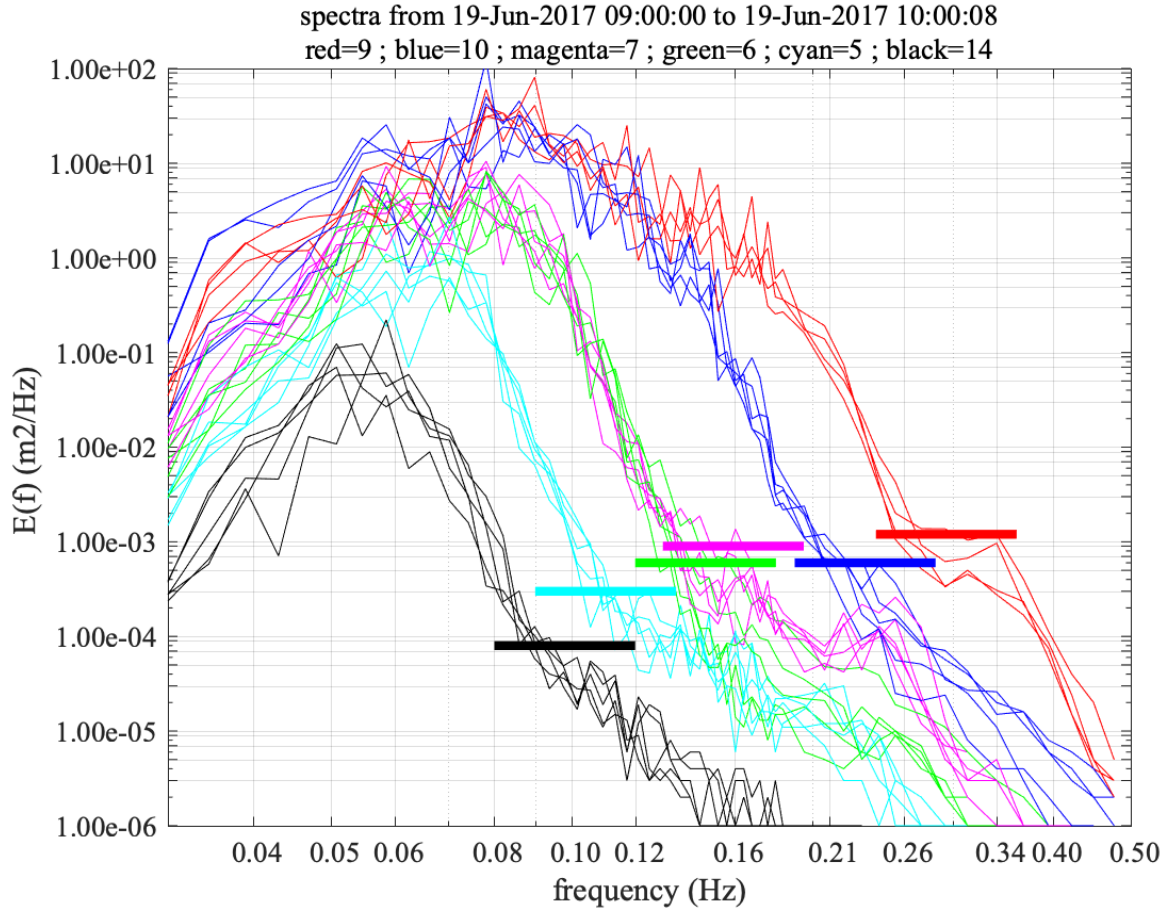

Figure 3. Energy spectra from the six buoys during the period of 0900 to 1000 UTC 19 June 2017. This uses the same color scheme as Figure 2. The thick horizontal lines indicate energy levels below which the tail is visibly "propped up", perhaps by instrument noise.

## 4. Wave model

### 4.1. General description

The wave model used in this study WAVEWATCH III® (WW3, Tolman 1991, WW3DG 2016). This is a phase-averaged spectral model for which the prognostic variable is wave action spectral density, which is the wave energy spectral density divided by the wave frequency: $N = E/\sigma$, where $\sigma = 2\pi/T$ ($T$ denoting wave period). The spectrum is a function of wavenumber or frequency ($k$ or $\sigma$), direction ($\theta$), space ($x$, $y$ or longitude, latitude), and time ($t$). The left hand side of the radiative transfer equation includes terms for time rate of change and propagation in the four dimensions (kinematics), while the right hand side includes source functions (dynamics):

$$\frac{\partial N}{\partial t} + \nabla \cdot \vec{c} N = \frac{S}{\sigma} \tag{1}$$

where $\vec{c}$ is a four-component vector describing the propagation velocities in $x$, $y$, $k$, and $\theta$. For example, in absence of currents, $c_x$ is the $x$-component of group velocity $C_g$. The sum of all source functions is denoted as $S$, and individual source functions are denoted with appropriate subscript: $S_{in}$, $S_{wc}$, $S_{nl4}$, and $S_{ice}$ being energy input from wind, dissipation by whitecapping, four-wave nonlinear interactions, and dissipation by sea ice, respectively. In deep water, without ice cover, the terms $S_{in}$, $S_{wc}$, and $S_{nl4}$ dominate $S$. We use the "source term package" of Ardhuin et al. (2010) known as 'ST4', for $S_{in}$ and $S_{wc}$. In this package, swell dissipation (weak losses of energy not associated with breaking) is formally part of $S_{in}$. For $S_{nl4}$, we use the Discrete Interaction Approximation (DIA) of Hasselmann et al. (1985). In finite water depth, other terms such as dissipation by bottom friction become important, but we omit these from the present discussion, as they are not expected to affect wave conditions in the dataset used here.

$S_{ice}$ is scaled by areal ice fraction $a_{ice}$, and the default behavior of WW3 is to scale open water source terms by the open water fraction, $1 - a_{ice}$:

$$S = (1 - a_{ice})(S_{in} + S_{wc} + S_{nl4}) + a_{ice}S_{ice} \qquad (2)$$

The scaling of open water source terms is an important concept in our analysis. WW3 permits the user to apply partial scaling, or omit the scaling entirely, and this can be done individually for each of the three open water terms. In case of full omission of the scaling for all three terms, we have:

$$S = S_{in} + S_{wc} + S_{nl4} + a_{ice}S_{ice} \qquad (3)$$

The scaling of $S_{in}$ is particularly debatable, since the transfer of energy from wind to waves occurs through normal stresses, and one can easily imagine that normal stresses remain effective (or partially effective) in cases where the ice is composed of frazil, brash, or pancakes, or when floe size is small relative to wave length. The scaling of $S_{nl4}$ is similarly in doubt. Some articles imply that non-scaling of $S_{nl4}$ is settled science, but support for this is thin: see discussion of relevant literature in Rogers et al. (2016). Scaling of $S_{wc}$, on the other hand, is unlikely to be as consequential. This is because 'ST4' uses a threshold-based breaking term, as proposed by Young and Babanin (2006). In the case of substantial ice cover, the spectrum will be sufficiently suppressed by $S_{ice}$ such that the model spectral density will tend to be below the breaking threshold (especially where $S_{ice}$ is strongest, say above 0.2 Hz), so $S_{wc}$ will tend to be small or zero, regardless of its scaling.

*Neither* representation (2) *nor* (3) are expected to be an optimal representation of what occurs in the real ocean. Rather, that optimal representation is somewhere in between, and probably is frequency-dependent. In particular, assumption (3) will give unrealistic results, e.g. permitting generation of large waves by a local wind event over continuous ice.

Where $k_i$ gives the exponential decay rate of amplitude in the space domain, the exponential decay rate of energy in the time domain, prior to scaling by $a_{ice}$, is computed as $D_{ice} \equiv S_{ice}/E = -2C_g k_i$. The group velocity $C_g$ can, in principle, be affected by ice cover, but observational support for this is primarily in the higher frequencies (Cheng et al. 2017; Collins et al. 2018), and theoretical models of the effect in the low and middle frequencies (say, $f < 0.25$ Hz) are inconsistent and even contradictory in outcome. In this study, we simply assume that the group velocity is the open water group velocity.

An important clarification must be made regarding the scaling of dissipation by ice concentration. In analysis of observational studies, where a positive correlation between dissipation rate and ice concentration is reported, it is crucial to note whether the authors are referring to $S_{ice}$ or $a_{ice}S_{ice}$. In the case of Kohout et al. (2020), the positive correlation is referring to the latter, scaled term. In Section 6.3, we look at correlation between ice concentration and the former, unscaled term (or, more precisely, $k_i$).

### 4.2. Implementation for this study

The specific implementation of WW3 used in this study is described in this section.

Wind forcing in the form of 10-m wind vectors comes from archives of the U.S. Navy's global atmospheric model, NAVGEM (Hogan et al. 2014), at 3 hourly intervals and 1/4˚ geographic resolution. Ice concentration forcing for the nested grid come from AMSR2 analyses provided by the University of Hamburg. This uses the "ARTIST" algorithm as described by Spreen et al. (2008) and Beitsch et al. (2013, 2014). This ice concentration product is at relatively high geographic resolution (median spacing is 3.05 km) but relatively low temporal resolution (one field every 24 hours).

The wave model grid receives boundary forcing from a global model hindcast. The global model was run 0000 UTC May 23 2017 to 0000 UTC June 30 2017. The global grid design known as "Irregular-Regular-Irregular" (Rogers and Linzell 2018) is used. Resolution is 1/4˚ at low latitudes and 18 km south of 50˚S.

Open water source terms and spectral grid settings used here are fairly typical of routine large-scale modeling using WW3. As noted already, the 'ST4' package for open water source terms is used with the 'DIA' method of computing nonlinear interactions. The spectral grid includes 36 directional bins and 31 frequency bins (0.0418 to 0.73 Hz, logarithmically spaced). The wind input source term of Ardhuin et al. (2010) requires specification of a parameter, $\beta_{max}$ which is used to compensate for the mean bias of the input wind fields, or lack thereof; $\beta_{max}$=1.2 is used for these hindcasts.

The nested grid bounds are indicated in Figure 1 and Figure 4. The PIPERS grid is bounded by 140˚E, 195˚E, 70.1˚S, and 60˚S and contains $292 \times 10^3$ sea points. The computational grids are polar stereographic and match the grids of the ice concentration input (3.05 km median resolution).

The northern limit of the nest, 60˚S, is selected so that all relevant ice is fully inside the nest during the simulation period. This implies that the $S_{ice}$ settings of the global model are inconsequential to the inversion.

The WW3 nest and example output is shown in Figure 4. The time period, 0900 June 19 UTC 2017, corresponds to the time period used in Figure 3. Buoy positions are marked with filled circles, using the color scheme of Figure 3. Buoy 9 (red circle) and 10 (blue circle) are used in later, in Figure 5. Note that for this case, the mean wave direction approximately matches the axis between these two buoys. An animation of images similar to Figure 4 are provided with the Supplemental Information.

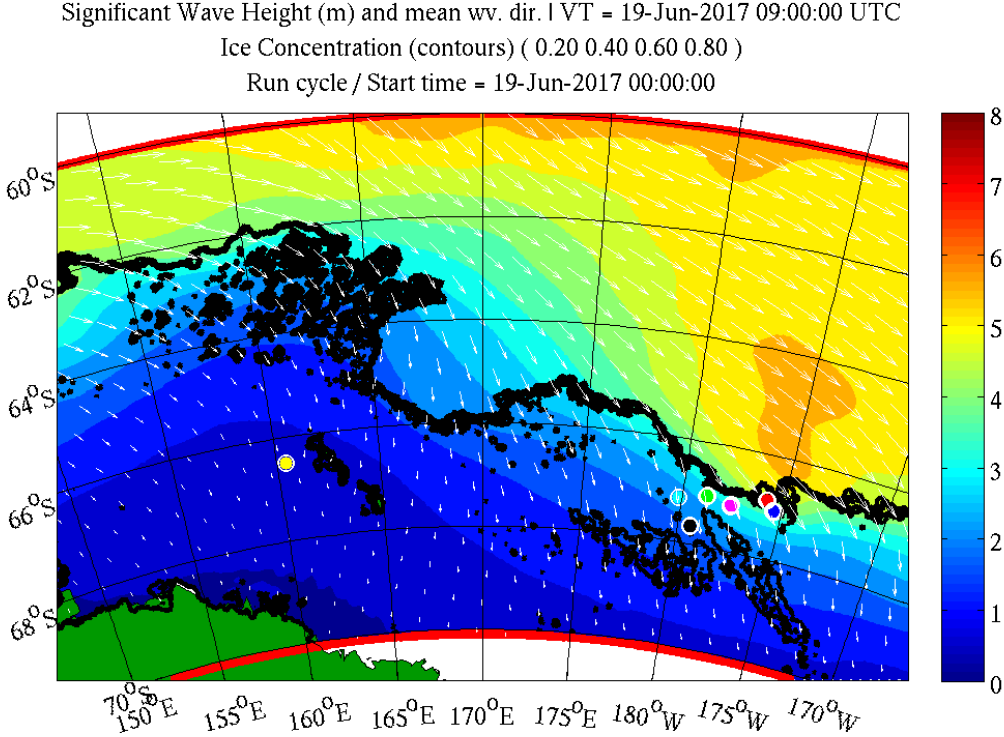

Significant Wave Height (m) and mean wv. dir. l VT = 19-Jun-2017 09:00:00 UTC
Ice Concentration (contours) ( 0.20 0.40 0.60 0.80 )
Run cycle / Start time = 19-Jun-2017 00:00:00

Figure 4. Wave conditions predicted by WW3 for the period of 0900 UTC, 19 June 2017. Red lines indicate the northern and southern boundary of the WW3 nest. Colors indicate the significant waveheight in meters. Arrows indicate mean wave direction. Contours indicate ice concentration at 20%, 40%, 60%, and 80%. This is the same time period used in Figure 3. Buoys are marked with dots, using the color scheme of the prior two figures. The yellow dot is the last surviving buoy in the western/inbound buoy group, not used in this paper.

## 5. Inversion process: introduction

Here we use the inversion method introduced in Rogers et al. (2016). In briefest terms, the objective is to determine the value of attenuation rate $k_i(f)$ which provides a match between the modeled energy density spectrum $E(f)$ and the corresponding observed $E(f)$. Thus, for each observational $E(f)$, a dissipation profile $k_i(f)$ is determined.

The discrete frequencies on which the modeled and observed $E(f)$ are natively described have dissimilar minimum, maximum, and spacing. For the purpose of the inversion, these are remapped to new distributions over the region in which they overlap, 0.042 to 0.472 Hz. These new frequency bins are intentionally coarser than either of the native frequency distributions, to increase degrees of freedom and improve stability of $E(f)$ estimates (e.g. Elgar 1987).

There are 16 coarse frequency bins, with spacing at or near 0.020 Hz at lower frequencies, and spacing at or near 0.039 Hz at higher frequencies. The left (i.e. low frequency) side of the first bin is 0.042 Hz. The right side of the second and last bin are 0.08 Hz and 0.472 Hz respectively. The number of bins is an increase relative to the eight frequency bins used in Rogers et al. (2016). In part, this is because of climatology of the Southern Ocean relative to the Beaufort Sea (the existence of energy at lower frequencies).

The primary computational component of the inversion consists of repeated simulations using the simplistic 'IC1' parameterization of WW3: each model run with IC1 uses a fixed value of $k_i$. Linear interpolation is performed between these values to find the "optimal" $k_i$, where observed $E(f)$ should match the modeled $E(f)$. The process is described and illustrated in Rogers et al. (2016).

The inversion includes 20 $k_i$ values tested using 20 hindcasts with IC1 (slightly more than Rogers et al. (2016), who used 17). The spacing is finer for the smaller values. In units of 1/m, they are: [0, 1e-6, 2e-6 ,..., 2000e-6 , 4000e-6 , 9000e-6]. Note that since negative values of $k_i$ are not included in this set, solutions of negative $k_i$ are effectively disallowed. This is discussed further below.

Since the global grid is not re-run for each of the 20 inversion members, the inversion requires that all relevant ice is represented within the computational grid of the IC1 hindcasts, i.e. wave energy does not pass through ice prior to reaching the nest boundary. This condition is satisfied in our case (Figure 4).

Since the inversion seeks to match the buoy observations by selecting the optimal $k_i$ value, and the source term $S_{ice}$ is scaled by ice concentration $a_{ice}$, it does not return a valid solution in cases of $a_{ice}$=0, and for cases of small $a_{ice}$, the solution $k_i$ is less stable. Therefore, these cases are omitted. We use the semi-arbitrary cut-off of $a_{ice}$=0.22. As noted in Section 3, we use data from the eastern buoy group collected during the period starting 0000 UTC 6 June and ending 0000 UTC 30 June 2017.

Here, we briefly list the advantages and disadvantages of the method relative to the traditional, geometric method:

1) (Advantage) With the inversion method, inclusion of non-ice source terms such as wind input is automatic. Typically, with the geometric method, these are omitted. The simple application of the geometric method (disregarding these other source terms), will suffer less from this omission when treating pure swell cases. We know of only one example where the non-ice source terms are included with the geometric method: Cheng et al. (2017). However, they used an early variant of the non-ice source terms (denoted 'ST1' in WW3).

2) (Disadvantage) The inversion method assumes that mismatch between observed and modeled $E(f)$ is caused by error in $k_i(f)$. The inversion results are therefore affected by numerical error (from finite differencing), errors in boundary forcing, errors in wind forcing, and errors in the non-ice source terms. The inversion process will seek to counter these errors using the $S_{ice}$ term, and so the $S_{ice}$ (or more precisely $k_i$) estimate is less accurate when these other errors are significant relative to $S_{ice}$.

3) (Neutral). Both methods are affected by any inaccuracy in the estimated ice concentration.

4) (Advantage). With the inversion method, one only needs one buoy to create a dissipation estimate. With the geometric method, at least two buoys are needed.

5) (Advantage). The geometric method disregards the time required for waves to travel from one buoy to another. If wave or ice conditions change during this time, it produces error. Thus, the geometric method becomes less accurate when there is larger spacing between buoys.

6) (Advantage). The geometric method is most accurate when the waves are travelling along the axis connecting the two buoys. In the case of the PIPERS deployment, this became an important advantage, because the buoys drifted over time to a non-optimal alignment (Section 3). The geometric method can be improved by accounting for directional spread and obliqueness, as done by Cheng et al. (2017), but this assumes uniformity of wave conditions perpendicular to the axis, so some error is introduced regardless.

7) (Disadvantage). The inversion method solves for $k_i$ at each frequency bin independently. Thus, it assumes that $k_i$ at one frequency bin does not affect wave spectra at other bins. Insofar as nonlinear interactions occur, this assumption introduces error.

8) (Advantage). The inversion method provides an estimate that is most relevant to the wave model. Insofar as the forward model and inverse model have similar errors (e.g. from numerics), the forward model which uses dissipation estimated from the application of inversion will benefit from some cancellation of errors.

9) (Disadvantage). The inversion method is much more difficult to apply than the simplest type of geometric estimate. However, the sophisticated geometric estimate of Cheng et al. (2017) also requires significant human effort.

10) (Neutral). The inversion method's dissipation rate corresponds to dissipation that occurs between the ice edge and the buoy. The geometric method's dissipation rate corresponds to dissipation that occurs between two buoys. In cases where both buoys are far from the ice edge, the two methods are estimating quite different quantities.

Above, we make a distinction between the <u>rigorous geometric method</u> used by Cheng et al. (2017) and the <u>simple geometric method</u> used in most (or all) other studies. Figure 5 compares the <u>inversion method</u> with the <u>simple geometric method</u> for a single time step. Colored lines are from the same PIPERS example used in prior figures (0600 to 1200 UTC 19 June 2017). Thin green lines are $k_i(f)$ profiles estimated by the inversion method at the inner buoy, which is #10 in this case. Thin red lines are $k_i(f)$ profiles estimated by the inversion method at the outer (near the ice edge) buoy, which is #9 in this case. There are multiple lines for each because the entire period from 0600 to 1200 UTC 19 June 2017 is used. The inversion estimates $k_i(f)$ at each buoy, independent of other buoys. The thick blue line is the estimate of $k_i(f)$ using the traditional geometric approach. The blue line corresponds to an average for the six-hour period. Three reference profiles are also shown in this figure. "SIPEX-12" is the fitted profile from Meylan et al. (2014). "SWIFT WA3" is the profile fitted to an inversion (Rogers et al. 2018b) using SWIFT buoys (Thomson 2012) in "Wave Array 3" of the Sea State field experiment (Thomson et al. 2018). The PIPERS "I1L" reference profile is an average profile from the present study, presented in Section 7 and Figure 8.

In the inversion, there is no guarantee that the dissipation of wave energy by sea ice will be much larger than the errors in the wave model. (This pertains to item 2 above.) In our case, this was most evident at the two lowest frequency bins. At low frequencies, $k_i$ is small, and its impact may be smaller than model errors, such as the error in the boundary forcing. For example, if our global model is providing boundary spectra that is too weak (low energy), the inversion will seek to compensate by reducing $k_i$. Since unmeasured true $k_i$ is already small, the inversion may even report that $k_i = 0$ is too large, implying negative dissipation. These solutions are treated as $k_i = 0$. In cases where these values are applied to create "mean $k_i$" values, we must be aware of the impact of this limiter $k_i = 0$: it will tend to shift the mean upward. For this reason, in our evaluation of mean profiles, we treat these first two bins with enhanced suspicion. These bins correspond to wave periods larger than 12.5 seconds.

At high frequencies, another challenge exists. When the buoys report zero energy, the inversion method is not applicable, since linear exponential decay cannot dissipate energy to zero. Also, the lower the energy density $E(f)$, the more likely it is to be substantially affected by noise (Figure 3). Thus, below a threshold level for energy density $E(f)$, the solution for that frequency, $k_i(f)$, is flagged as unusable for this study. The threshold is somewhat subjective, and a higher threshold reduces the population of solutions at high frequencies. Here, we conservatively assume that the change in the slope of the high-frequency tails in Figure 3 is caused by noise, and apply an algorithm to flag the noise-contaminated values. Since noise increases with total energy, a threshold based on observed significant waveheight is used rather than a fixed $E(f)$ threshold. The algorithm is described in Appendix A. The outcome of including a less restrictive threshold (so more $k_i(f)$ solutions) is discussed in Section 8.

As noted above, an error is introduced to the inversion results when there is model-observation mismatch that is not associated with $S_{ice}$. For the wave frequency range of 0.08 to 0.472 Hz, we expect that the treatment of the "open water source terms" ($S_{in}$, $S_{ds}$, and $S_{nl4}$) in grid cells containing ice is a source of uncertainty. We have performed two inversions to bound the uncertainty:

- Inversion 1: open water source terms are scaled with open water fraction (the default in WW3). This is presented in the present manuscript.
- Inversion 2: open water source terms are not directly affected by ice cover. For sake of brevity, this is presented in the Supplemental Information.

As noted already in Section 4.1, the unknown correct scaling is probably somewhere in between, e.g. wind input can likely occur through ice, but only for certain ice conditions and wavelengths.

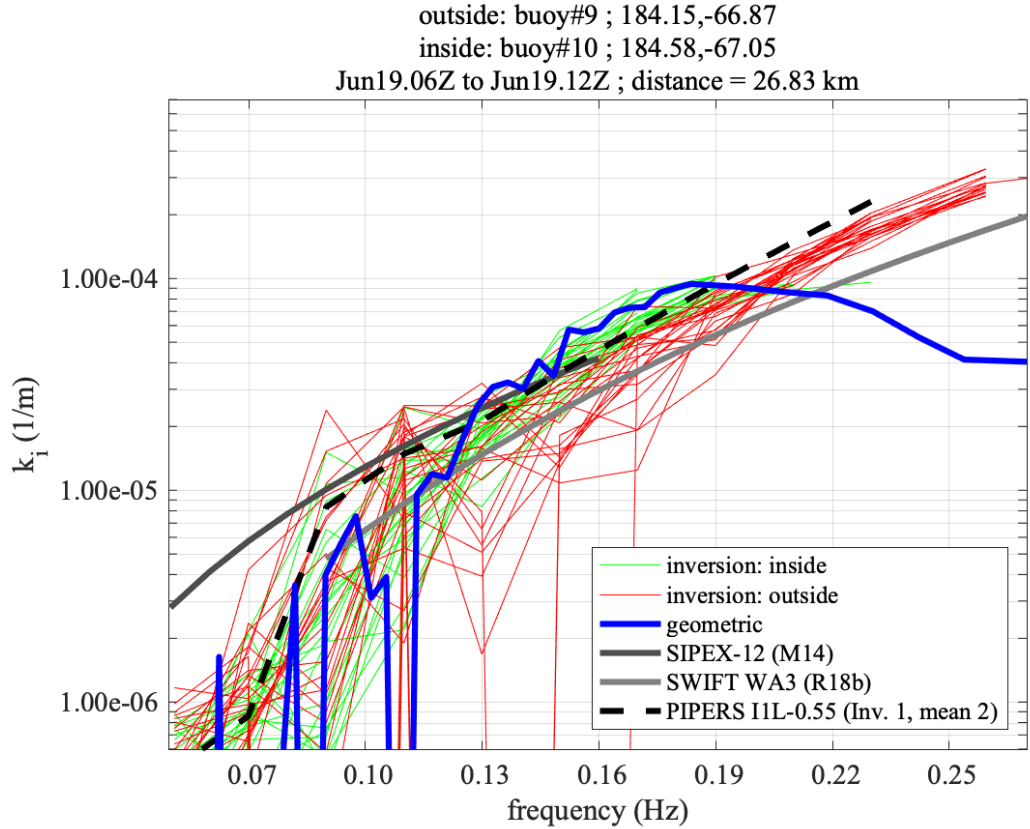

Figure 5. Estimated dissipation rates, as a function of frequency ("dissipation profiles", $k_i(f)$), comparing the inversion results (thin green and red lines) against the results from the simple geometric method (thick blue line). Grey and black lines are reference profiles. See text for explanation.

## 6. Inversion results (dissipation profiles)

This section discusses the $k_i(f)$ profiles resulting from the application of the inverse method to the PIPERS dataset. The overall goal is to identify the correlation between the $k_i(f)$ profiles and other variables. Since $k_i$ and frequency are the primary and secondary variables presented, these are referred to here as "tertiary variables".

Figure 6 illustrates this concept and explains the approach used in subsequent plots. Here, the tertiary variable is arbitrary and is denoted as $R$. The approach is to bin the $k_i(f)$ profiles of a large dataset according to $R$ and indicate $R$ as in Figure 6, using the color scaling. We can imagine at least three different scenarios:

1) In this scenario (illustrated in Figure 6), though there is considerable variability in the $k_i(f)$ profiles, they are well sorted by $R$. This implies that if $R$ is predictable, then $k_i(f)$ is predictable.

2) In a second scenario, not illustrated, all of the $k_i(f)$ profiles lie nearly on top of each other (i.e. small variability). This scenario implies again that $k_i(f)$ is predictable, but this time, prediction of $R$ is unnecessary.

3)  In the third scenario, also not illustrated, there is considerable variability in the $k_i(f)$ profiles and there is not apparent sorting by (i.e. correlation with) $R$. This scenario that even with knowledge of $R$, $k_i(f)$ cannot be predicted well.

What would be the best candidates for a tertiary variable? We can imagine that the condition or character of the ice cover is the primary determiner of $k_i(f)$. This is the concept used by Rogers et al. (2016): they used photos taken from buoy-mounted cameras to characterize the ice cover and demonstrated sorting of $k_i(f)$ by this variable. However, this "ice condition" was not general: it only applied to the frazil, brash, and pancake ice encountered during that field experiment. More importantly, to use such a quantity in a numerical wave model, the ice condition would need to be available as a gridded variable, and this is well beyond present technology. In this paper, we use tertiary variables that are available from modeling or remote sensing. These are listed in Section 6.1.

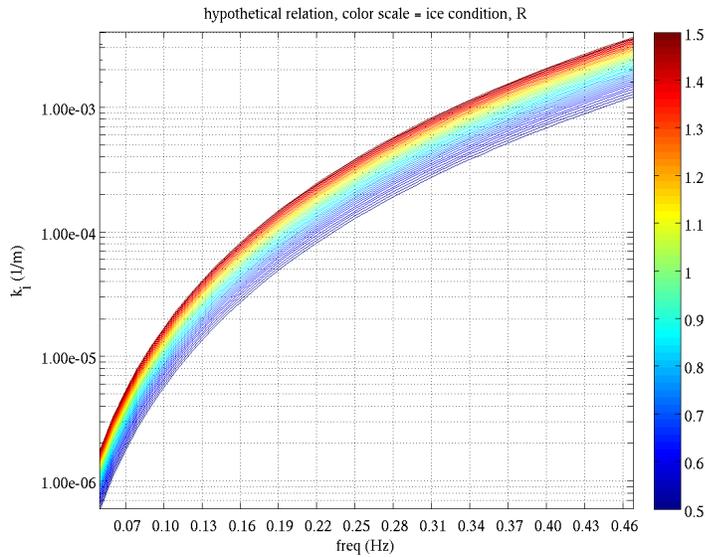

Figure 6. Hypothetical set of dissipation profiles. See text for explanation.

## 6.1. Methods: co-located variables (tertiary parameters)

In this section, we list the tertiary variables which we will inspect for correlation with $k_i(f)$ profiles from the inversions.

Ice concentration, $a_{ice}$ estimates are taken from AMSR2 and are the same fields as used in the model forcing (Section 4.2).

Ice thickness, $h_{ice}$ estimates are derived from the MIRAS radiometer onboard the European Space Agency's SMOS satellite. Processed files are provided by the Univ. Bremen and downloaded from https://seaice.uni-bremen.de/thin-ice-thickness/ . These files are on a 12 km polar stereographic grid, with one analysis per day. Ice thickness ($h_{ice}$) values are available at ice thicknesses up to 50 cm, where the instrument saturates. Therefore, cases of $h_{ice}$=50 cm presented here should be interpreted as $h_{ice}$>=50 cm. In cases where the daily field is missing, the $h_{ice}$ value is flagged as unusable. The reader is referred to Huntemann et al. (2014) and Paţilea et al. (2019) for more information about the SMOS "thin sea ice" thickness retrieval.

Ice distance, $x_{ice}$ is computed as the distance from the buoy to the ice edge. Since the buoys are in ice, the algorithm searches for the nearest open water point ($a_{ice} = 0$). Isolated open water points are disregarded. This estimate is approximate insofar as it is not necessarily the distance along the main axis of wave propagation.

Significant waveheight, $H_{m0}$. This parameter is taken from buoy observations. $H_{m0}$ is computed as $H_{m0} = 4\sqrt{m_0}$, where $m_n = \int E(f)f^n df$, and $E(f)$ is the one dimensional (non-directional) energy spectrum, and the bounds of integration are defined by the overlap with model spectral range, i.e. 0.042 to 0.476 Hz.

Seven other tertiary variables are included in the Supplemental Information and Mendeley Data archive: mean period, $T_m$; spectral fourth moment, $m_4$; significant steepness, $S_{sig}$; representative orbital velocity, $V$; wind speed, $U_{10}$; air temperature; and date.

Our purpose here is to identify correlations. These correlations may or may not indicate causal relations, i.e. direct dependence of $k_i(f)$ on the tertiary variable. A number of these variables are computed from the wave spectrum itself ($H_{m0}$, $T_m$, $m_4$, $S_{sig}$, and $V$). If the relation is causal, this implies *nonlinearity* of the dissipation. Such behavior would not be extraordinary, e.g. whitecapping dissipation $S_{wc}$ is *extremely* nonlinear, and in fact, we already use a few nonlinear forms of $S_{ice}$ in WW3, e.g. the dissipation by ice-water friction under orbital motion as implemented by F. Ardhuin in Stopa et al. (2016, Appendix B) and some of the empirical forms implemented by Collins and Rogers (2017). Extensive discussion of potential nonlinearity of $S_{ice}$ can be found in Squire (2018).

## 6.2. Methods: Quantifying role of tertiary parameters

We use inversion results from the nine buoys in the eastern group (see Section 3) during the period of June 6 to June 30, and only consider cases with $a_{ice}$>0.22. This results in 9477 valid $k_i$ profiles. In the cases where $h_{ice}$ is the tertiary parameter, the population is reduced further (to 8957), to only include $k_i$ profiles with contemporaneous $h_{ice}$ estimates, see Section 6.1.

The $k_i(f)$ profiles are binned by the value of the tertiary variables, with 15 to 20 bins, depending on the tertiary variable. Bin widths are determined such that they have a nominally equal population of $k_i(f)$ profiles. These are used to create the mean $k_i(f)$ profiles which are plotted, with coloration according to the mean of the tertiary parameter within those cases. We have 9477 valid $k_i(f)$ profiles, so for example with 20 bins, there would 474 $k_i(f)$ profiles represented in each bin, except for the last bin which has 471. As with the other binning method, averaging is performed separately for each frequency bin, since the population of $k_i(f)$ also varies by frequency. Within each bin, the mean $k_i(f)$ profile is treated as unreliable for a given frequency $f$ (i.e. not plotted) if less than 24 values (5% of the maximum) are used in the averaging.

## 6.3. Results

Both binning methods are applied in Figure 7 to the four tertiary variables ($h_{ice}$, $a_{ice}$, $x_{ice}$, $H_{m0}$). Similar comparisons for other tertiary parameters and Inversion 2 can be found in the Supplemental Information. The gray rectangle in each panel indicates the first two frequency bins of the inversion. As noted in Section 5, the inversion results for this frequency range are potentially most affected by errors in boundary forcing.

Some noteworthy features of Figure 7 are:

1) The trend for ice concentration, $a_{ice}$ for $f$>0.10 Hz, there is more dissipation with lower ice concentration. This may seem counterintuitive, but keep in mind that $k_i$ plotted here is the dissipation of only the ice-covered fraction of the sea. We scale $S_{ice}$ by ice concentration in the model, $S_{ice} = -2a_{ice}C_g k_i E$, and the inversion provides the $k_i$ prior to the scaling. The behavior is likely caused by coincidental relations: we see smaller $k_i(f)$ occurring further from the ice edge (large $x_{ice}$).

2) At low and mid-frequencies ($f$<0.13 Hz), there is a clear increase in $k_i(f)$ with $h_{ice}$. This is an intuitively causal relation. This result is consistent with a primary conclusion of Robin (1963), who observed that ice thickness is the most useful parameter to predict the penetration of waves into sea ice for lower frequencies ($f$<0.10 Hz). At higher frequencies, the trend is either absent or reversed, because there is a noticeable flattening of the $k_i(f)$ profiles for thicker ice. This result is likely spurious; the flattening is discussed in Section 8.

3) Unsurprisingly, the dependence on $x_{ice}$ is broadly similar to the dependence on $h_{ice}$. (There is thicker ice further from the ice edge.)

4) There is a clear trend of smaller $k_i(f)$ for larger waveheight, $H_{m0}$. A possible causal relation would be the breakup of floes during large events, making the ice cover more pliable and less dissipative. This is a nonlinear type of dissipation. Considering the ice cover as quasi-viscous in aggregate, this is a thixotropic behavior (shear thinning) and is reversed from the more typical nonlinearity for wave model dissipation terms, e.g. higher dissipation for larger orbital velocity, or larger steepness, etc. The fact that the correlation is primarily in the energy-containing region of the spectrum is consistent with this causality. This correlation is consistent with the results of Montiel et al. (2018), who find that for larger waveheights ($H_{m0}$> 3 m), $k_i$ is inversely proportional to $H_{m0}$ (i.e. linear, non-exponential decay).

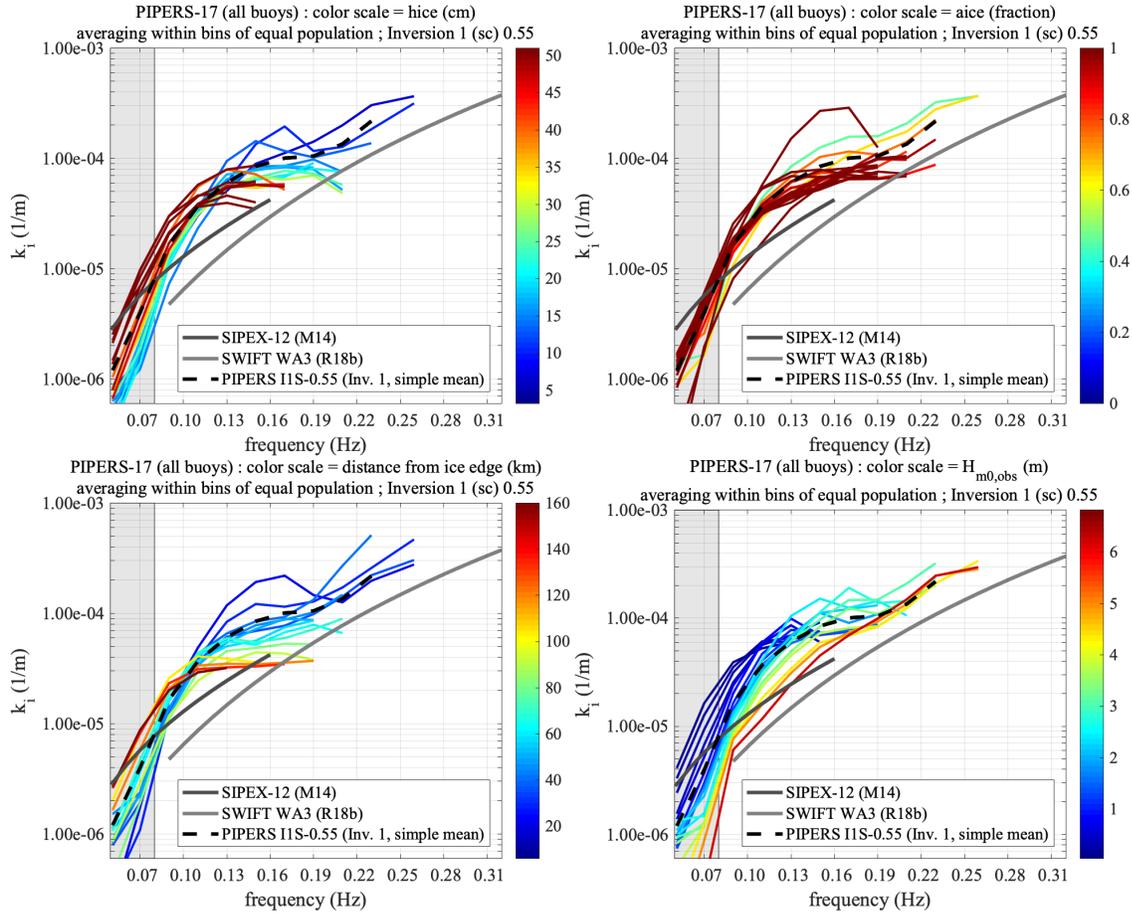

Figure 7. Dissipation profiles estimated using the model-data inversion. Four variables are evaluated here: ice thickness, ice concentration, distance from ice edge, and significant waveheight. The "SIPEX" and "SWIFT" reference profiles (gray and black lines) are introduced in Section 5. The PIPERS "I1S" is a simple mean profile from the present study (see Section 7).

## 7. Fitting to parametric model: two examples

Empirically-derived mean $k_i$ profiles can be fitted to a parametric form. Several possible forms exist, but here, we look at binomial subsets of the general polynomial $k_i = \sum C_n f^n$. This follows the approach of Meylan et al. (2014), who used $C_2 f^2 + C_4 f^4$, i.e. $n = 2$ and 4. The values of $n$ can be interpreted as indicators of a dominant physical process (or two processes in the case of a binomial) if they match the $n$ predicted by a theoretical model. For example, if an empirical $k_i$ profile fits well to $n = 7$, this would support the argument that dissipation is caused by something like viscosity in the ice layer (Keller model, see Section 2). The theories of Section 2 primarily support integer values of $n$, with one exception ($n = 3.5$).

Herein, we use two methods of creating mean profiles. Both stop at the 10th (of 16) frequency bins, since the number of valid $k_i(f)$ values beyond 0.23 Hz is relatively small. The two methods are:

1) A "simple mean" profile, denoted "I1S". As long as a $k_i(f)$ value is considered as valid, it is included in the average. As with the binned profiles in Figure 7, the mean profile is terminated when the population of the frequency bin is less than 5% of the maximum, which is at 0.23 Hz. As an example, this averaging includes 9464, 4738, and 524 $k_i$ values at bins centered at 0.09, 0.15, and 0.23 Hz respectively.

2) A mean using only the profiles which span the full frequency space of the mean profile, from 0.05 and 0.23 Hz in terms of bin centers. This averaging includes 487 $k_i$ values at all of these bins, and thus provides greater consistency at the cost of reduced population. This averaging is denoted "I1L" ("L" for "long profiles").

Most invalid $k_i$ values are marked as such because $E(f)$ falls below the noise estimate (Appendix). Method I1L thus primarily censors cases of thicker ice further from the ice edge (large $h_{ice}$ and $x_{ice}$), and omits most profiles exhibiting "flattening" at higher frequencies which is likely spurious (Section 8).

There are many possible mean profiles that could be created from this dataset[2]. Though not done here, it is possible to use a weighted mean $k_i$ profile as the basis for a fit. For example, we could use the fit with $k_i$ weighted by waveheight to some power if we wanted to emphasize $S_{ice}$ dissipation under high wave conditions.

Figure 8 shows fits to mean profile I1S (left-side panels) and I1L (right-side panels). The top panels show the mean $k_i$ profiles on log-log scale, so that the slope of the line segments corresponds to values of $n$. Integer values of $n$ are indicated as straight lines on these panels. We see that I1S has slope corresponding to $n = 2$ for frequencies 0.11 to 0.21 Hz, but the overall slope is somewhat steeper. Mean profile I1L has mean slope corresponding to $n = 2$ for the frequencies 0.09 to 0.13 Hz and close to $n = 4$ for the overlapping range 0.11 to 0.23 Hz. Based on this, and goodness-of-fit for various tested binomials (not shown), we use binomial fits of $n = 2$ and 3 for I1S and $n = 2$ and 4 for I1L. The power dependence of the latter is consistent with prior analyses using different datasets: see the empirical fits of Meylan et al. (2014) and Rogers et al. (2018b).

Middle panels of Figure 8 illustrate the error minimization process used to find the best fit values of $C_n$. The diagonal error distribution in both cases suggests that the dependence is fairly evenly distributed between the two terms of each binomial. Best fit coefficients for I1S are $C_2$=1.82e-3 ; $C_3$=8.70e-3, and for I1L, they are $C_2$=4.10e-4 ; $C_4$=6.09e-2. In both examples, the first two bins are excluded during the fitting process, since inversion results are more sensitive to errors in boundary forcing for these bins.

Dependence on $n = 2$ to 4 can be found in a number of theoretical models reviewed in Section 2, namely: two new theoretical models of Meylan et al. (2018) ($n = 2$ and $n = 3$), the turbulent boundary layer model ($n = 3$), the Robinson and Palmer (1990) model ($n = 3$), the inextensible surface cover model ($n = 3.5$), and the Kohout et al. (2011) friction model ($n = 4$).

The fitted polynomials are shown in the lower panels. The quality of fit for I1S is less good, owing to the flattening near the center (0.15 to 0.19 Hz). The quality of fit of I1L is excellent. In the lower right panel, the I1L binomial is compared to earlier binomial fits, from Meylan et al. (2014) and Rogers et al. (2018b). The dissipation estimates from the former, for broken floes near Antarctica is largely consistent with I1L, while estimates from the former, for pancake and frazil ice near the Beaufort Sea, are lower than I1L, by factors of 1.5 to 1.8.

---

[2] Readers interested in doing this themselves may download the dissipation profiles and co-located tertiary variables from Mendeley Data, http://dx.doi.org/10.17632/5b742jv7t5.1 .

*Limitations*

Because in the underlying dataset there is variance about the mean profiles, any simple parametric model that is only a function of frequency will have a limited skill, even when directly applied to the same case: the predicted $E(f)$ will also match the observed in the mean, but with significant variance. Inclusion of tertiary variables, such as $h_{ice}$, in the parametric formula is a logical next step.

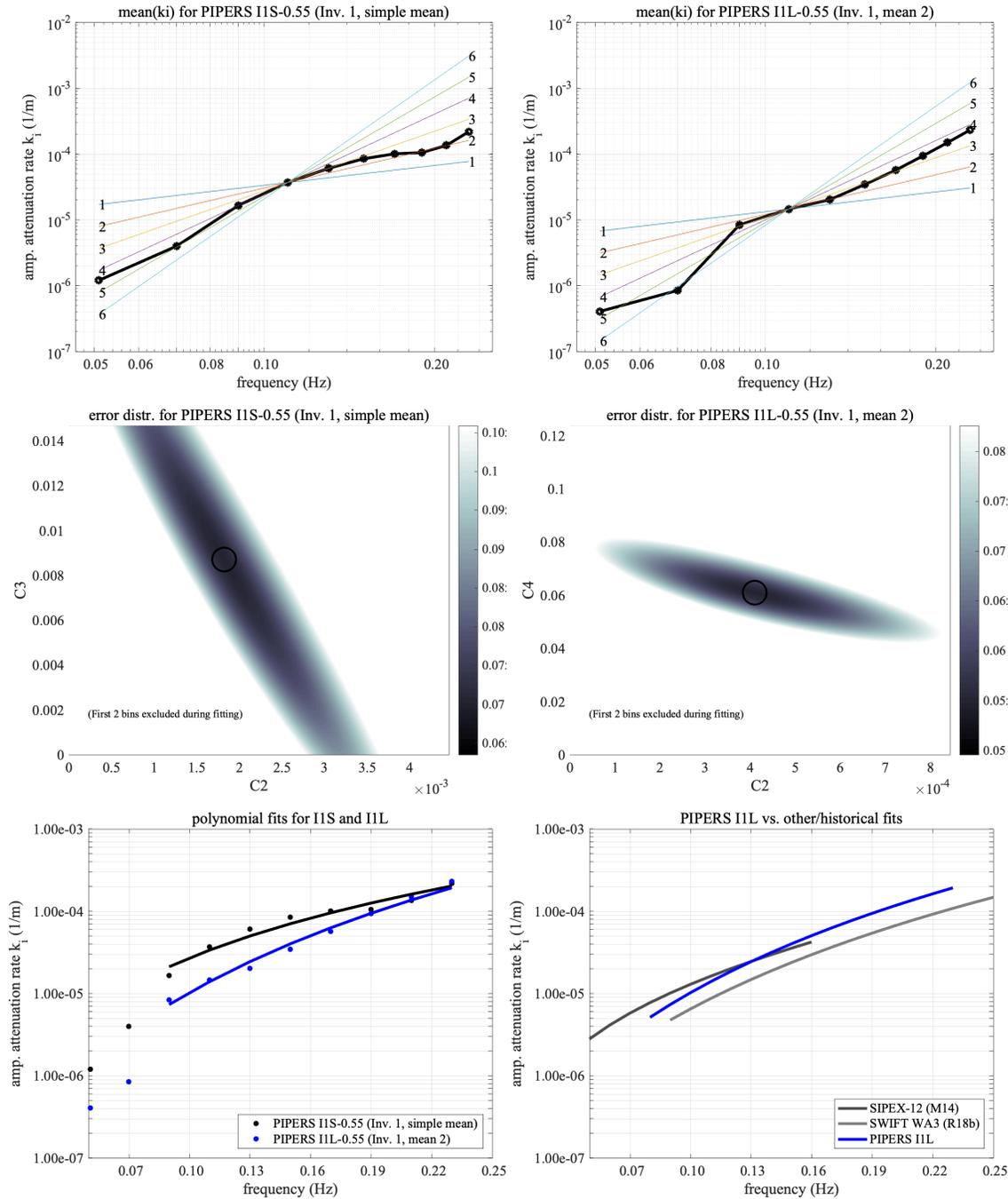

Figure 8. Top: Evaluation of power dependence of $k_i$ on frequency. (Power dependence corresponds to slope on this log-log scale.) Integers within axes indicate power dependence. Middle: Error surface for determining best fit coefficients. Left: from mean profile (I1S). Right: from mean profile (I1L). Bottom left: Best fit binomials for I1S and I1L (note that horizontal axis is linear scale). Bottom right: Comparison of I1L binomial to historical binomials. See text for further explanation.

## 8. Discussion

*Flattening of dissipation profiles*

With some of the binned $k_i$ profiles, there is a flattening of the profiles, either at mid-frequencies or at both mid- and high frequencies. This flattening is more evident with Inversion 1 presented here, which scales the open water source terms with open water fraction, than in Inversion 2, presented in the Supplemental Information, which does not scale these source terms. Flattening is most noticeable for low values of $H_{m0}$ and high values of $h_{ice}$ and $x_{ice}$.

This flattening can be thought of as a nascent or more subtle variation on the "roll-over" effect noticed in observational data by Wadhams et al. (1988) and others. In fact, Wadhams et al. (1988) themselves offer the possibility that this non-monotonic behavior is an artifact of non-representation of wind input, as they use the simple geometric method to compute $k_i(f)$. With Inversion 1, though wind input is represented, it may be under-represented insofar as the scaling by open water fraction may represent too large a reduction of wind input. Thus, it is not surprising that Inversion 1 has this subtler form of the roll-over.

However, we acknowledge that this is not the only explanation. We can imagine several possibilities, of which the above belongs in the group (2b):
1. The flattening of the profiles may, in fact, represent the correct form of dissipation.
2. The flattening may be artificial and associated with:
   a) Errors in the inversion not associated with source terms, e.g. boundary forcing, numerics, or wind and ice input fields.
   b) Incorrect representation of included model source terms, such as an overprediction of the reduction of wind input by partial ice cover.
   c) Source terms which exist in nature but are absent in the model. Specifically, a source term that would tend to "prop up" the middle and/or high frequencies of the spectrum. For example, scattering and reflection (e.g. Shapiro and Simpson 1953) are not represented in the model. Also, it is possible that the motion of floes induced by the dominant waves creates secondary motions of the free surface.
   d) As discussed in Section 3, instrument noise would also tend to "prop up" the high-frequency portion of the spectrum. This would tend to affect low-waveheight cases the most, consistent with the behavior seen in the $k_i(f)$ profiles in Figure 7. Though we aggressively flag suspect $E(f)$ values as unusable based on $E(f)$ inspection (see Appendix), noise in $E(f)$ can affect $k_i(f)$ estimates even when the noise is too small to be apparent in $E(f)$ plots (J. Thomson, manuscript in preparation).

Items 2b and 2c would each imply that the Inversion's $k_i(f)$ profiles are not presenting pure dissipation, but some mix of dissipation and "generation" at the affected frequencies. Further, if any flattening of $k_i(f)$ in the present study is assumed to be spurious, this helps to explain counterintuitive outcomes, e.g. Items 2c or 2d may explain the negative correlation between ice thickness and dissipation at 0.13 to 0.16 Hz in Figure 7.

*Minimum energy threshold*

As noted in Section 5, we apply a threshold on energy density, below which the $k_i(f)$ solution is flagged as unusable due to possible contamination by noise. That threshold, described in the Appendix, is relatively conservative (high) and depends on the total energy of each spectrum. We also experimented with a more liberal (lower) threshold, $E(f) = 1 \times 10^{-5}$ m²/Hz, or 10 times the precision of the dataset. The primary outcome is that the $k_i(f)$ profiles extend to much higher frequencies, primarily 0.40 to 0.45 Hz. There are only minor differences in conclusions. The best fit slopes are unchanged ($n = 2,3$ for Inversion 1), and goodness of fit is similar, with modest impact on the coefficients of proportionality.

*Indeterminable causation*

With this dataset, we are only able to quantify correlations and then make educated guesses about the causal relations. In some cases, the causations may be indirect: variable $X$ affects variable $Y$, which then affects variable $Z$, so $X$ and $Z$ correlate. For example, higher wave conditions may make the ice cover more pliable, which in turn reduces dissipation. Also, dissipation correlates with distance from the ice edge, but dissipation rate likely depends on ice conditions, while ice conditions are more directly related to distance from ice edge. In other cases, the correlation may be coincidental, variables $X$ and $Y$ both have a causal dependence on variable $Z$, so $X$ and $Y$ correlate. Lastly, there may be cases where variables probably affect each other simultaneously, as is likely the case with the wave conditions and the (unmeasured) ice conditions. So-called "process models", with two-way interactions and resolution of individual floes and wave phase, may be the best approach for studying such relations. However, there will always be uncertainty from underlying assumptions and missing processes.

## 9. Summary and Conclusions

In this section, we summarize the study, list the conclusions based on the analysis presented, and lastly, we present thoughts and speculations about the results.

*Summary*

- A new dataset is presented: $k_i(f)$ profiles (dissipation rate as a function of frequency) from model/data inversion. These are computed using observations of waves in ice in the Southern Ocean presented by Kohout et al. (2020). 24 days (June 6 to 30 2017) are used, which is a subset of a more extended dataset.
- The $k_i(f)$ profiles are evaluated by studying their correlation with 11 environmental variables (denoted "tertiary variables") such as ice thickness $h_{ice}$ and significant waveheight $H_{m0}$. This multi-dimensional analysis is made possible by the large population of the dataset (9477 spectra).
- The $k_i(f)$ profiles are presented using a novel methods of binning by tertiary parameters.
- Parametric models of dissipation are created by fitting to mean $k_i(f)$ profiles.

*Conclusions*

- At lower frequencies ($f < 0.12$ Hz), there is a clear increase of $k_i(f)$ with $h_{ice}$.
- At low and mid-frequencies ($f < 0.20$ Hz), wave energy parameters such as waveheight, $H_{m0}$, negatively correlate with dissipation rate.

- The dissipation rate is negatively correlated with $a_{ice}$, $h_{ice}$, and $x_{ice}$ at middle and higher frequencies ($f > 0.13$ Hz). This is presumed to be a spurious result, likely caused by artificial flattening of profiles for cases further from the ice edge.
- The simple mean of $k_i(f)$ profiles, denoted "I1S", exhibits a flattening in middle frequencies, 0.15 to 0.19 Hz. The fitting to a binomial form $n = 2$ and 3 is fair, where $k_i(f) = \sum C_n f^n$.
- A second mean of $k_i(f)$ profiles, denoted "I1L" is created by excluding $k_i(f)$ profiles that terminate early ($f < 0.23$ Hz). Thus the profile is dominated by cases of thinner ice closer to the ice edge, and less sensitive to negative effects of instrument noise. This profile is well described by a binomial using $n = 2$ and 4.
- Of the two fits, the binomial for I1L is qualitatively more similar to the binomials from Meylan et al. (2014) and Rogers et al. (2018). The former was computed for a case of broken floes in Antarctica using simple geometric calculations, and is roughly comparable to I1L. The latter was computed using an inversion with wind input scaling, for an example of pancake and frazil ice in the western Arctic, and these values are generally lower than those of I1L, by factors 1.5 to 1.8.

*Further discussion*

- Power dependence found here ($n = 2$ and 3) and ($n = 2$ and 4) can be compared with our review of power dependence in the literature (Section 2). For example, $n = 3$ dependence is consistent with an existing model for dissipation caused by turbulence at the ice-water interface, as well as two other theoretical models, described in Meylan et al. (2018).
- Variability of the $k_i(f)$ profiles about the simple mean profiles is significant. If these mean profiles are applied as a parametric dissipation rate in a forward model for the same application, results can be expected to be accurate only in the mean sense. Future incorporation of tertiary variables such as $h_{ice}$ or $H_{m0}$ in the parametric model may improve model skill.
- The negative correlation between dissipation rate and energy follows the analog to a thixotropic (shear-thinning) ice cover. This is a loose analogy insofar as we do not advocate general treatment of the ice layer as a viscous fluid ($n = 7$).
- The physical causation associated with the tertiary variables is not proven here. For example, correlations may be associated with common dependence on ice conditions which are not measured.
- Positive correlation between ice thickness and dissipation rate, found here, may be exploited in development of parametric dissipation formula and applied in predictive models.
- The "correct" scaling, i.e. that which best matches the unknown, correct source terms in the real ocean is probably somewhere in between the two treatments of this study (Inversion 1 in this manuscript and Inversion 2 in the Supplemental Information), and is perhaps frequency-dependent.

**Declaration of Competing Interest**

The authors declare that they have no known competing financial interests or personal relationships that could influence nor appear to influence the work reported in this paper.

**Acknowledgments**

Author ER was funded by the Office of Naval Research via the NRL Core Program, Program Element Number 61153N. The 6.2 project was titled "Wave-ice interactions". Author MM was funded by the Australian Research Council, grant DP200102828. Author AK was funded by New Zealand's Deep South National Science Challenge Targeted Observation and Process-Informed Modelling of Antarctic Sea Ice and NIWA core funding under the National Climate Centre Climate Systems programme.

We thank the crew of the R/V Nathaniel B. Palmer and the members of the science team on the PIPERS crew who helped with the buoy deployment and ice observations. We thank the University of Hamburg for providing the AMSR2 ice concentration analysis. We thank the University of Bremen and Dr. Li Li (NRL Remote Sensing Division) for providing the satellite-based ice thickness estimates. We thank Mr. Michael Phelps (NRL Oceanography Division contractor) for providing archives of the NAVGEM fields. We thank Drs. David Wang (NRL), Jim Thomson (UW/APL), and Tripp Collins (USACE) for helpful information and advice about instrument error and data processing.

This is NRL contribution number NRL/JA/7320-20-4950 and is approved for public release.

**Appendix: Noise threshold**

An algorithm was developed for this study, to remove $E(f)$ records which may contain a significant proportion of noise.

Traditionally, instrument noise most severely affects the measurement of long, low-amplitude waves. For example, if the acceleration spectrum measured by a buoy is contaminated by noise that is uniform in frequency (white noise), then the contamination of $E(f)$ will fall off at a slope of $f^{-4}$, and tend to be extremely small at higher frequencies, relative to an open water spectrum. However, in sea ice, this is not necessarily the case, and as noted in our discussion of Figure 3, the high-frequency tail appears to be "propped up", and this may be caused by noise. We conservatively assume that this feature *is* caused by noise.

Instrument noise is expected to increase with total energy, so our approach is designed to include this dependency. The development was initiated by sorting spectra by significant waveheight $H_{m0}$ and putting in groups of 50, so the first group has 50 spectra with the lowest $H_{m0}$, and so on. The groups were then manually evaluated to estimate the $E(f)$ below which the spectra appear to be contaminated, similar to Figure 3. In a second phase, an objective method was created, following these steps:

1) For each group of 50 spectra, a mean spectrum was computed. The averaging was performed in log-space.
2) The slope of the mean spectrum was computed, in log-log space.
3) The point in frequency space where the negative slope in the tail was steepest was identified. This is assumed to be the point where dissipation by sea ice is very strongly suppressing the tail while noise is negligible.
4) The point in frequency space where the magnitude of slope reaches less than 55% of the maximum is identified. This is treated as the threshold $T$, where if $E(f)$ is below the $T$, that $E(f)$ bin should be flagged as unusable. The criterion of 55% was determined as the setting which most closely matches the subjective (manual) evaluation.

5) Since the spectra were sorted by $H_{m0}$ prior to grouping, the computed thresholds were then available as $T = T(\langle H_{m0}\rangle)$, where $\langle H_{m0}\rangle$ is the mean $H_{m0}$ of each group. We then performed a least squares fit linear regression to $T(\langle H_{m0}\rangle)$ in log-log space, resulting in a power fit, $T = 3.99 \times 10^{-4} H^{0.85}$.

Since a different accelerometer was used for buoys 13 and 14, we perform a separate analysis for these buoys, giving $T = 3.89 \times 10^{-3} H^{1.86}$. These represent only 550 out of 9477 of the spectra, so a smaller group size (20) was used for these buoys. The two fits are illustrated in Figure A1.

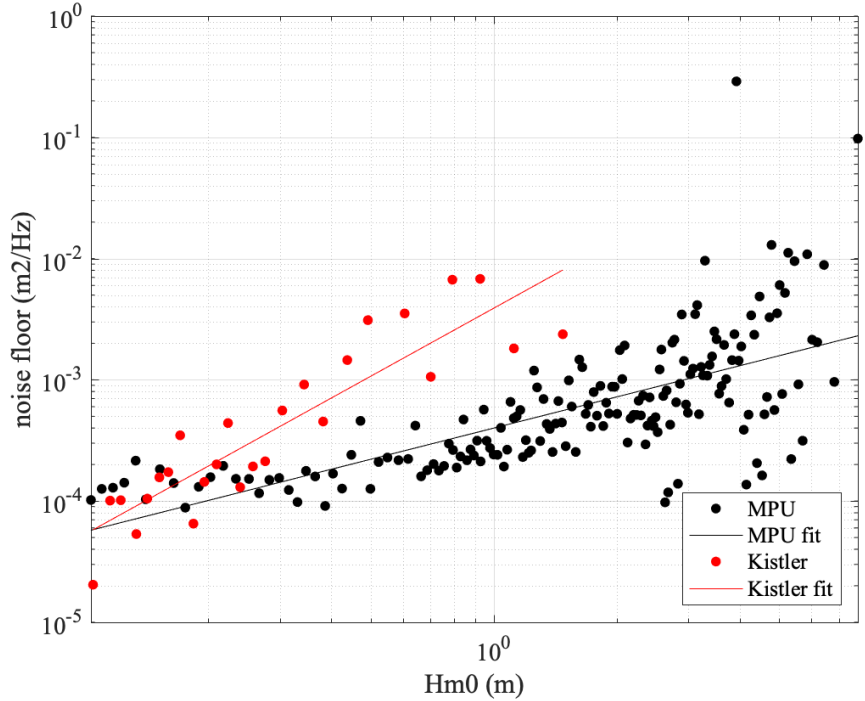

Figure A1. Noise threshold $T$ as a function of waveheight. Dots indicate results using the algorithm based on spectral slope. Lines indicate the power fits applied to flag $E(f)$ values as unusable in this study. "MPU" and "Kistler" denote two types of accelerometers used. "MPU" accounts for 94.2% of the measured spectra, and "Kistler" represents the other 5.8% (buoys 13 and 14).

**Data Availability**

The PIPERS wave spectra can be downloaded from
ftp://ftp.niwa.co.nz/incoming/Kohout/WII_2017 . Dissipation profiles presented here are available from Mendeley Data, http://dx.doi.org/10.17632/5b742jv7t5.1. Other data files used to create figures herein are available from the corresponding author by request.

**References**

Alberello, A., Onorato, M., Bennetts, L., Vichi, M., Eayrs, C., MacHutchon, K., and Toffoli, A., 2019. Brief communication: Pancake ice floe size distribution during the winter expansion of the Antarctic marginal ice zone, *The Cryosphere*, **13**, 41–48, https://doi.org/10.5194/tc-13-41-2019.

Ardhuin, F., Rogers, W. E., Babanin, A.V., Filipot, J.-F., Magne, R., Roland, A., van der Westhuysen, A., Queffeulou, P., Lefevre, J.-M., Aouf, L. and Collard, F., 2010. Semi-empirical dissipation source functions for ocean waves: Part I, definitions, calibration and validations, *J. Phys. Oceanogr.* **40**, 1917-1941.

Ardhuin, F., P. Sutherland, M. Doble, and P. Wadhams, 2016. Ocean waves across the Arctic: Attenuation due to dissipation dominates over scattering for periods longer than 19 s, *Geophys. Res. Lett.,* **43**, 5775–5783, doi:10.1002/2016GL068204.

Asplin, M.G.; J. Marko, D.B. Fissel, K. Borg, 2018. Investigating Propagation of Short-Period Ocean Waves Into the Periphery of Arctic Pack Ice Using High-Resolution Upward-Looking Sonar, *Atmosphere-Ocean*, 56 (3), 152-161, https://doi.org/10.1080/07055900.2018.1498765.

Beitsch, A., L. Kaleschke, S. Kern, and X. Tian-Kunze, 2013. AMSR2 ASI 3.125 km Sea Ice Concentration Data, V0.1, Institute of Oceanography, University of Hamburg, Germany, digital media. (ftp-projects.zmaw.de/seaice/).

Beitsch, A., L. Kaleschke, and S. Kern, 2014. Investigating High-Resolution AMSR2 Sea Ice Concentrations during the February 2013 Fracture Event in the Beaufort Sea, *Remote Sens.*, **6**, 3841-3856

Bennetts, L.G., Williams, T.D., 2015. Water wave transmission by an array of floating discs. *Proc. R. Soc. A*, 471:20140698

Bennetts L.G., O'Farrell, S., Uotila, P., 2017. Brief communication: impacts of ocean-wave-induced breakup of Antarctic sea ice via thermodynamics in a stand-alone version of the CICE sea-ice model, *Cryosphere*, 11:1035–40

Bidlot, J., 2018. Intercomparison of operational wave forecasting systems against buoys: September to November 2017, Report for The Joint WMO-IOC Technical Commission for Oceanography and Marine Meteorology (JCOMM), 90 pp., retrieved from http://www.jcomm.info/wave on 8 April 2020.

Boutin, G., Ardhuin, F., Dumont, D., Svigny, C., Girard-Ardhuin, F., Accensi, M., 2018. Floe size effect on wave-ice interactions: possible effects, implementation in wave model, and evaluation. *J. Geophys. Res. Oceans* **123**, 4779–805

Cavaleri, L., S. Abdalla, A. Benetazzo, L. Bertotti, J.R. Bidlot, Ø. Breivik, S. Carniel, R.E. Jensen, J.P. Yandum, W.E. Rogers, A. Roland, A. Sanchez- Arcilla, J.M. Smith, J. Staneva, Y. Toledo, G.Ph. van Vledder, and A.J. van der Westhuysen, 2018: Wave modelling in coastal and inner seas. *Progress in Oceanography,* **167,** 164-233, https://doi.org/10.1016/j.pocean.2018.03.010.

Cheng, S., W.E. Rogers, J. Thomson, M. Smith, M.J. Doble, P. Wadhams, A.L. Kohout, B. Lund, O.P.G. Persson, C.O. Collins III, S.F. Ackley, F. Montiel and H.H. Shen, 2017. Calibrating a Viscoelastic Sea Ice Model for Wave Propagation in the Arctic Fall Marginal Ice Zone. *J. Geophys. Res.,* **122** doi://10.1002/2017JC013275

Collins, C.O., and W.E. Rogers, 2017. A Source Term for Wave Attenuation by Sea ice in WAVEWATCH III®: IC4, *NRL Report NRL/MR/7320--17-9726*, 25 pp. [available from www7320.nrlssc.navy.mil/pubs.php].

Collins, C., M. Doble, B. Lund, and M. Smith, 2018. Observations of surface wave dispersion in the marginal ice zone. *J. Geophys. Res.,* **123**. doi:10.1029/2018JC01378.

Doble, M. J., 2009. Simulating pancake and frazil ice growth in the Weddell Sea: A process model from freezing to consolidation, *J. Geophys. Res.*, **114**, C09003, doi:10.1029/2008JC004935.

Doble, M. J., and P. Wadhams, 2006. Dynamical contrasts between pancake and pack ice, investigated with a drifting buoy array, *J. Geophys. Res.*, 111, C11S24, doi:10.1029/2005JC003320.

Doble, M. J. and Bidlot, J.-R., 2013. Wavebuoy measurements at the Antarctic sea ice edge compared with an enhanced ECMWF WAM: progress towards global waves-in-ice modeling, *Ocean Model.*, **70**, 166-173, doi:10.1016/j.ocemod.2013.05.012.

Doble, M. J., De Carolis, G., Meylan, M.H., Bidlot, J.-R. and Wadhams, P., 2015. Relating wave attenuation to pancake ice thickness, using field measurements and model results, *Geophys. Res. Lett.*, 42, 4473-4481, doi:10.1002/2015GL063628.

Elgar, S., 1987. Bias of effective degrees of freedom of a spectrum, *J. Waterway, Port, Coastal, and Ocean Eng.*, **113**(1)*,* 77-82.

Gemmrich, J., W.E. Rogers, J. Thomson, and S. Lehner, 2018: Wave evolution in off-ice wind conditions *J. Geophys. Res.: Oceans,* **123**, 5543–5556, https://doi.org/10.1029/2018JC013793.

Hasselmann, S., Hasselmann, K., Allender, J.H. and Barnett, T.P., 1985. Computations and parameterizations of the nonlinear energy transfer in a gravity-wave spectrum. Part II: Parameterizations of the nonlinear energy transfer for application in wave models. *J. Phys. Oceanogr.*, 15, 1378-1391.

Herman, A., Cheng, S., and Shen, H.H., 2019. Wave energy attenuation in fields of colliding ice floes – Part 1: Discrete-element modelling of dissipation due to ice–water drag, *The Cryosphere*, 13, 2887–2900, https://doi.org/10.5194/tc-13-2887-2019.

Hogan, T., and 16 Coauthors, 2014. The Navy Global Environmental Model. *Oceanogr.,* **27**(3), 116-125.

Huntemann, M., Heygster, G., Kaleschke, L., Krumpen, T., Mäkynen, M., and Drusch, M., 2014. Empirical sea ice thickness retrieval during the freeze-up period from SMOS high incident angle observations, *The Cryosphere*, **8**, 439-451, doi:10.5194/tc-8-439-2014.

Keller, J.B.,1998. Gravity waves on ice-covered water. *J. Geophy. Res.*, **103** (C4): 7663-7669.

Kohout, A.L., Williams, M., 2019. Antarctic wave-ice observations during PIPERS. NIWA client report 2019060CH prepared for the Deep South Challenge. National Institute of Water and 470 Atmospheric Research, New Zealand. Contact: library@niwa.co.nz.

Kohout, A.L., M.H. Meylan, D.R. Plew, 2011. Wave attenuation in a marginal ice zone due to the bottom roughness of ice floes. *Annals of Glaciology*, **52**(57), 118-122.

Kohout, A. L., M. J. M. Williams, S. M. Dean, and M. H. Meylan, 2014. Storm-induced sea-ice breakup and the implications for ice extent, *Nature,* **509***,* 604–607.

Kohout, A. L., B. Penrose, S. Penrose, M.J.M. Williams, 2015. A device for measuring wave-induced motion of ice floes in the Antarctic marginal ice zone. *Annals of Glaciology,* **56** (69), 415-424.

Kohout, A.L., M. Smith, L.A. Roach, G. Williams, F. Montiel, M.J.M. Williams, 2020. Observations of exponential wave attenuation in Antarctic sea ice during the PIPERS campaign, submitted to *Annals of Glaciology.*

Lamb, H., 1932. *Hydrodynamics*. Sixth Edition. New York, Dover Publications, 768 pp.

Li, J., Kohout, A.L., and Shen, H.H., 2015a. Comparison of wave propagation through ice covers in calm and storm conditions, *Geophy. Res. Lett.* **42**(14)*,* 5935–5941, doi: 10.1002/2015GL064715.

Li, J., S. Mondal, H.H. Shen, 2015b. Sensitivity analysis of a viscoelastic parameterization for gravity wave dispersion in ice covered seas. *Cold Reg. Sci. Technol.* **120**, 63–75.

Li, J., Kohout, A. L., Doble, M. J., Wadhams, P., Guan, C., and Shen, H. H., 2017. Rollover of apparent wave attenuation in ice covered seas. *J. Geophy. Res.*, **122**. https://doi.org/10.1002/2017JC012978.

Liu, A.K. and E. Mollo-Christensen, 1988. Wave propagation in a solid ice pack. *J. Phys. Oceanogr.*, **18**, 1702-1712.

Liu, A. K., B. Holt, and P. W. Vachon, 1991. Wave propagation in the Marginal Ice Zone: Model predictions and comparisons with buoy and Synthetic Aperture Radar data. *J. Geophys. Res.*, **96**, (C3), 4605-4621.

Liu, Q., Rogers, W. E., Babanin, A. V., Young, I. R., Romero, L., Zieger, S., Qiao, F. and Guan, C., 2019. Observation-based source terms in the third-generation wave model WAVEWATCH III: updates and verification, *J. Phys. Oceanogr.* 49. 489-517.

Liu, Q., Rogers, W.E., Babanin, A., Li, J., and Guan, C., 2020. Spectral modelling of ice-induced wave decay, *J. Phys. Oceanogr.*, https://doi.org/10.1175/JPO-D-19-0187.1

Marchenko, A., Wadhams, P., Collins, C., Rabault, J., Chumakov, M., 2019. Wave-ice interaction in the north-west barents sea. *Appl. Ocean Res.* 90, 101861. ISSN 0141-1187. https://doi.org/10.1016/j.apor.2019.101861.

Meylan, M., L.G. Bennetts, and A.L. Kohout, 2014. In situ measurements and analysis of ocean waves in the Antarctic marginal ice zone, *Geophys. Res. Lett.*, **41**, 5046–5051, doi:10.1002/2014GL060809.

Meylan, M.H., Bennetts, L.G., Mosig, J.E.M., Rogers, W.E., Doble, M.J. and Peter, M.A., 2018. Dispersion relations, power laws, and energy loss for waves in the marginal ice zone, *J. Geophys. Res.*, 123, 3322-3335. https://doi.org/10.1002/2018JC013776.

Montiel F, Squire V.A., Bennetts L.G., 2016. Attenuation and directional spreading of ocean wave spectra in the marginal ice zone. *J. Fluid Mech.* **790**, 492–522.

Mosig, J.E.M., F. Montiel, and V.A. Squire, 2015. Comparison of viscoelastic-type models for ocean wave attenuation in ice-covered seas, *J. Geophys. Res.*, **120**, 6072–6090, doi:10.1002/2015JC010881.

Montiel, F., Squire, V.A., Doble, M.J., Thomson, J., Wadhams, P., 2018. Attenuation and directional spreading of ocean waves during a storm event in the autumn Beaufort Sea marginal ice zone. *J. Geophys. Res.*, **123**, 5912–32

Orzech, M., Shi, F., Veeramony, J., Bateman, S., Calantoni, J., and Kirby, J., 2016. Incorporating floating surface objects into a fully dispersive surface wave model, *Ocean Model.*, 102, 14–26, https://doi.org/10.1016/j.ocemod.2016.04.007.

Paţilea, C., Heygster, G., Huntemann, M., and Spreen, G., 2019. Combined SMAP–SMOS thin sea ice thickness retrieval, *The Cryosphere*, 13, 675–691, https://doi.org/10.5194/tc-13-675-2019.

Rabault, J., Sutherland, G., Jensen, A., Christensen, K.H., Marchenko, A., 2019. Experiments on wave propagation in grease ice: Combined wave gauges and PIV measurements. *J. Fluid Mech.* 864, 876–898. https://doi.org/10.1017/jfm.2019.16

Roach L.A., Smith, M.M., Dean S.M., 2018. Quantifying growth of pancake sea ice floes using images from drifting buoys. *J. Geophys. Res. Oceans* 123:2851–66.

Robin, G.Q., 1963. Wave propagation through fields of pack ice. *Philosophical Transactions of the Royal Society of London. Series A, Mathematical and Physical Sciences*, **255**(1057) (Feb. 21, 1963), 313-339.

Robinson, N. J., and S. C. Palmer, 1990. A modal analysis of a rectangular plate floating on an incompressible liquid, *J. Sound Vibration*, **142**(3), 453–460, doi:10.1016/0022-460X(90)90661-I.

Rogers, W.E. and R.S. Linzell, 2018. The IRI Grid System for Use with WAVEWATCH III. *NRL Report NRL/MR/7320--18-9835*, 47 pp. [available from www7320.nrlssc.navy.mil/pubs.php]

Rogers, W. E. and M. D. Orzech, 2013. Implementation and testing of ice and mud source functions in WAVEWATCH III®. *NRL Memorandum Report, NRL/MR/7320-13-9462*, 31pp. [available from www7320.nrlssc.navy.mil/pubs.php].

Rogers, W.E., J. Thomson, H.H. Shen, M.J. Doble, P. Wadhams and S. Cheng, 2016. Dissipation of wind waves by pancake and frazil ice in the autumn Beaufort Sea, *J. Geophys. Res.*, **121,** 7991-8007, doi:10.1002/2016JC012251.

Rogers, W.E., P. Posey, L. Li, R. A. Allard, 2018a. Forecasting and hindcasting waves in and near the marginal ice zone: Wave modeling and the ONR "Sea State" field experiment. *NRL Report NRL/MR/7320--18-9786*, 179 pp. [available from www7320.nrlssc.navy.mil/pubs.php].

Rogers, W.E., M.H. Meylan, A.L. Kohout, 2018b. Frequency distribution of dissipation of energy of ocean waves by sea ice using data from Wave Array 3 of the ONR "Sea State" field experiment. *NRL Report NRL/MR/7322--18-9801*, 25 pp. [available from www7320.nrlssc.navy.mil/pubs.php].

Shapiro, A. and Simpson, L., 1953. The effect of a broken ice field on water waves, *Eos Trans. Am. Geophys. Union*, **34**, 36-42, https://doi.org/10.1029/TR034i001p00036

Shen, H.H., 2019. Modelling ocean waves in ice-covered seas, *Applied Ocean Res.*, **83**, doi:10.1016/j.apor.2018.12.009.

Shi, Y.F., Yang, Y.Z., Teng, Y., Sun, M., Shengjun, Y., 2019. Mechanism of sea ice formation based on comprehensive observation data in Liaodong Bay, China. *J. of Oceanol. and Limn.*, 37 (6): 1846-1856, https://doi.org/10.1007/s00343- 019-8269-8.

Spreen, G., L. Kaleschke, G. Heygster, 2008. Sea Ice Remote Sensing Using AMSR-E 89 GHz Channels, *J. Geophys. Res.*, **113**, C02S03, doi:10.1029/2005JC003384.

Squire, V.A., 1998. "The Marginal Ice Zone", in Physics of Ice-covered Seas, Vol. 1, edited by Matti Lepparanta, Helsinki University Printing House, Helsinki, 381-446.

Squire VA., 2018. A fresh look at how ocean waves and sea ice interact. *Philos. Trans. R. Soc. A* 376:20170342, 13 pp.

Squire, V. A., 2020. Ocean wave interactions with sea ice: a reappraisal, *Annual Rev. Fluid Mech.* 52, 37–60.

Sree, D.K.K., A.W.-K. Law, H.H. Shen, 2018. An experimental study on gravity waves through a floating viscoelastic cover, *Cold Reg. Sci. Technol.*, https://doi.org/10.1016/j.coldregions.2018.08.01

Steele, K.E. J.C. Lau, and Y.L. Hsu, 1985. Theory and application of calibration techniques for an NDBC directional wave measurements buoy. *IEEE Journ. Oceanic Eng.*, **10(4),** 382-396.

Stopa, J.E., F. Ardhuin, F. Girard-Ardhuin, 2016. Wave climate in the Arctic 1992-2014: seasonality and trends, *The Cryosphere*, **10**, 1605-1629.

Stopa, J. E., Sutherland, P., and Ardhuin, F., 2018. Strong and highly variable push of ocean waves on Southern Ocean sea ice, P. Natl. Acad. Sci. USA, 115, 5861–5865.

Sutherland, G., Rabault, J., Christensen, K.H., Jensen, A., 2019. A two layer model for wave dissipation in sea ice. *Appl. Ocean Res.* 88, 111–118. ISSN 0141-1187. https://doi.org/10.1016/j.apor.2019.03.023.

Sutherland, P., J. Brozena, W.E. Rogers, M. Doble and P. Wadhams, 2018. Airborne remote sensing of wave propagation in the marginal ice zone *J. of Geophys. Res.: Oceans*, 123, 4132-4152, https://doi.org/10.1029/2018JC013785.

Thomson, J., 2012. Wave breaking dissipation observed with "SWIFT" drifters, *J. Atmos. Ocean. Tech.,* **29**, 1866–1882, doi:10.1175/JTECH-D-12-00018.1.

Thomson, J., Talbert J., de Klerk, A., Brown, A., Schwendeman M., Goldsmith, J., Thomas, P., Olfe, C., Cameron, G. and Meinig, C., 2015. Biofouling effects on the response of a wave measurement buoy in deep water. *J. Atm. Ocean. Tech.*, 32, 1281-1286, doi: 10.1175/JTECH-D-15-0029.1.

Thomson, J. and 32 co-authors, 2018. Overview of the Arctic Sea State and Boundary Layer Physics Program, *J. Geophys. Res.*, 8674-8687, doi:10.1002/2018JC013766.

Toffoli, A., Bennetts, L.G., Meylan, M.H., Cavaliere, C., Alberello, A., Elsnab, J., Monty, J. P., 2015. Sea ice floes dissipate the energy of steep ocean waves. *Geophys. Res. Lett.* 42:8547–54.

Tolman, H.L., 1991. A Third-generation model for wind-waves on slowly varying, unsteady, and inhomogeneous depths and currents, *J. Phys. Oceanogr.* **21**(*6),* 782-797.

Vichi, M., Eayrs, C., Alberello, A., Bekker, A., Bennetts, L., Holland, D., et al., 2019. Effects of an explosive polar cyclone crossing the Antarcticmarginal ice zone. Geophysical Research Letters, 46. https://doi.org/10.1029/2019GL082457

Voermans, J., Babanin, A., Thomson, J., Smith, M., and Shen, H., 2019. Wave attenuation by sea ice turbulence, *Geophys. Res. Lett.*, 46, 6796–6803, https://doi.org/10.1029/2019GL082945.

Wadhams, P., 1978. Wave decay in the marginal ice zone measured from a submarine. *Deep-Sea Res.*, 25, 23-40.

Wadhams, P., Squire, V.A., Goodman, D.J., Cowan A.M. and Moore, S.C., 1988. The attenuation rates of ocean waves in the marginal ice zone, *J. of Geophys. Res.*, 93, 6799-6818.

WAVEWATCH III® Development Group (WW3DG), 2016. User manual and system documentation of WAVEWATCH III® version 5.16. *Tech. Note 329*, NOAA/NWS/NCEP/MMAB, College Park, MD, USA, 326 pp. + Appendices.

Weber, J.E., 1987. Wave attenuation and wave drift in the marginal ice zone. *J. Phys. Oceanogr.*, **17**(12), 2351-2361.

Worby, A. P. (1999), Observing Antarctic Sea Ice: A practical guide for conducting sea ice observations from vessels operating in the Antarctic pack ice. A CD-ROM produced for the Antarctic Sea Ice Processes and Climate (ASPeCt) program of the Scientific Committee for Antarctic Research (SCAR) Global Change (GLOCHANT) program, Hobart, Tasmania, Australia. See also: http://aspect.antarctica.gov.au/ , (retrieved 6 April 2020)

Young, I.R., and A.V. Babanin, 2006. Spectral distribution of energy dissipation of wind-generated waves due to dominant wave breaking. *J. Phys. Oceanogr.*, **36**, 376–394.

Yue C., Li, J.K., Guan, C.L., Lian, X.H., Wu, K.J., 2019. Surface wave simulation during winter with sea ice in the Bohai Sea. *J. of Oceanol. and Limn.*, 37 (6): 1 857-1 867, https://doi.org/10.1007/s00343- 019-8253-3.

Zhao, X., Cheng, S., Shen, H.H., 2017. Nature of wave modes in a coupled viscoelastic layer over water. *J. Eng. Mech.* 143:04017114.